\documentclass{mn2e}
\usepackage[pass]{geometry}
\usepackage{graphicx}
\usepackage{times}

\def\arcdeg{\hbox{$^\circ$}}
\def\arcmin{\hbox{$^\prime$}}
\def\arcsec{\hbox{$^{\prime\prime}$}}
\def\deg2{\hbox{$\rm deg^{2}$}}

\def\lsim{\mathrel{\rlap{\lower4pt\hbox{\hskip1pt$\sim$}}\raise1pt\hbox{$<$}}} 
\def\gsim{\mathrel{\rlap{\lower4pt\hbox{\hskip1pt$\sim$}}\raise1pt\hbox{$>$}}} 

\begin{document}
\title[Outbursts]{Results of a systematic search for outburst events in 1.4 million galaxies}
\author[A.J. Drake et al.]{
A.J.~Drake,$^1$ S.G.~Djorgovski,$^1$ M.J.~Graham,$^1$ D.~Stern,$^2$ A.A.~Mahabal,$^1$ 
\newauthor M.~Catelan,$^{3,4}$ E.~Christensen$^5$ and S.~Larson$^5$
\\
$^1$California Institute of Technology, 1200 E. California Blvd, CA 91225, USA\\
$^2$Jet Propulsion Laboratory, California Institute of Technology, 4800 Oak Grove Drive, MS 169-221, Pasadena, CA 91109, USA\\
$^3$Pontificia Universidad Cat\'olica de Chile, Instituto de Astrof\'isica, 
Facultad de F\'{i}sica, Av. Vicu\~na Mackena 4860, 782-0436 Macul, Santiago, Chile\\
$^4$Millennium Institute of Astrophysics, Santiago, Chile\\
$^5$The University of Arizona, Department of Planetary Sciences,  Lunar and Planetary Laboratory, 
1629 E. University Blvd, Tucson AZ 85721, USA\\
}

\volume{000}
\pubyear{0000}
\maketitle

\begin{abstract}
\noindent
We present an analysis of nine years of Catalina Surveys optical photometry
for 1.4 million spectroscopically confirmed SDSS galaxies. We find
717 outburst events that were not reported by ongoing transient
surveys. These events have timescales ranging from weeks to years. More than
two thirds of these new events are found in starforming galaxies, while such
galaxies only constitute $\sim20\%$ of our sample. Based on the properties
of the hosts and events, we find that almost all of the new events are
likely to be associated with regular supernovae. However, a small number of
long-timescale events are found among the galaxies containing AGN. These
events have similar properties to those recently found in the analyses of
light curves of large samples of AGN.  Given the lack of such events among
the more than a million passive galaxies in the sample, we suggest that the
long outbursts are associated with super-massive black holes or their
environments.

\end{abstract}

\section{Introduction}

Our understanding of the nature and range of variability in galaxies has
evolved significantly over the past decade. These advances are largely due
to the long-term monitoring of millions of galaxies by wide-field transient
surveys, such as the Catalina Real-time Transient Survey (CRTS, Drake et
al.~2009, Djorgovski et al.~2012), the Panoramic Survey Telescope and Rapid Response System
(PS1, Chambers et al.~2016), and the Palomar Transient Factory (PTF, Law et al.~2009).

The most common cause for rapid, transient variations in the brightness of galaxies
is supernovae. These objects provide insight into the final stages of stellar 
evolution, and, in the case of type-Ia supernovae, provide uniform
standard candles for measuring cosmological distances (Sandage and Tammann 1982).
To date, more than 25,000 supernova candidates have been 
discovered\footnote{http://www.rochesterastronomy.org/snimages/}.

In the past five years supernova discovery rates have climbed from hundreds
to thousands of events per year. This increase is almost exclusively due to
the advent of modern transient surveys. This increase has been so
dramatic that it far outstrips our ability to spectroscopically confirm the
nature of most transient discoveries.  Even with new spectroscopic surveys, such as
the Public ESO Spectroscopic Survey for Transient Objects (PESSTO; Smartt et al.~2015), 
less than 20\% of new supernova candidates are confirmed.  This situation has 
become known as the ``follow-up problem".

Aside from supernovae, it is well known that many galaxies harbor 
massive black holes that undergo accretion processes that are often 
observed as nuclear variability. The variations of these sources are 
usually a few tenths of a magnitude, but can become compounded to 
variations of more than a magnitude on long timescales (MacLeod et al.~2012,
Graham et al.~in prep.). 
In very rare cases, these Active Galactic Nuclei (AGN) have recently 
been found that appear to turn-on or off over timescales of years 
(MacLeod et al.~2016, Gezari et al.~2017, Stern et al.~2018.).
This so called "changing-look" type behaviour have been observed 
at low levels in target studies of Seyfert galaxies for decades 
(e.g. Tohline \& Osterbrock 1976; Osterbrock \& Shuder 1982; Goodrich 1989),
but has only recently been observed in luminous quasars. The advent of 
large surveys has confirmed this as a wide spread phenomenon.

A moderately large number (72 as of Jan.~2018)\footnote{https://tde.space/}
of transient events have also been associated with the tidal disruption of 
stars by massive nuclear black holes (Rees 1988, Gezari et al.~2003, 
Auchettl et al.~2017). 
Still further events temporal variability associated with AGN continue to 
have an unclear origin (e.g. Meusinger et al.~2010; Drake et al. 2011a). 
The possible causes of these AGN-related events are believed 
to include microlensing events, stellar mass black hole mergers, 
and superluminous supernovae (Lawrence et al.~2016, Graham et al.~2017).

The increased photometric monitoring of millions of galaxies by transient 
surveys has resulted in a significant increase in the number of transients 
found in galaxies. This increase has resulted in a smaller fraction of 
transients being spectroscopically confirmed. This is a trend that will almost 
certainly continue with even larger transient surveys, such as the Zwicky Transient 
Facility (ZTF; Bellm 2014) and the Large Synoptic Survey Telescope (LSST; 
Ivezic et al.~2008). The fact that we are unable to spectroscopically 
confirm the nature of most transients greatly increases the importance 
of any prior information that can be garnered from existing data.

Archival information from large spectroscopic surveys, such as the Sloan
Digital Sky Survey (SDSS; Albareti et al.~2017), can aid our understanding
of the nature of current and past transient events. For example, the 
spectroscopic redshift of a galaxy gives us a distance estimate for events 
that appear associated with it. The combination of this
distance with extinction values and a lightcurve enable us to estimate
the energetics of an event.  Furthermore, the presence of emission lines 
in the host spectrum can be used to estimate the star formation rate and
metallicity of the host.
In turn, these pieces of information can often be used to constrain the type 
of event.  For example, type-Ia supernovae have a well constrained set of 
lightcurve shapes and are found in all types of galaxies. In contrast, 
core collapse supernovae (hereafter CCSNe) are generally believed to be the 
product of massive stars with short lifetimes and are observationally very 
rare in ellipticals (van den Bergh \& Tammann 1991).

In this paper we investigate what can be learnt about the variability
in galaxies by combining existing photometry and spectra for more than a
million galaxies. We specifically search for significant flares and
outburst events that may have been missed in large surveys due to selection
biases or assumptions about galaxy variability. We investigate the general 
nature of the galaxies hosting significant outbursts.  We pay particular 
attention to the presence of flaring events like those found by Lawrence et al.~(2016) 
and Graham et al.~(2017) in AGN. Finally, we consider how archival data may 
help mitigate the follow-up problem in current and future large transient and 
variability surveys. We adopt $H_{0}\rm = 72\,\,km\,\,s^{-1} Mpc ^{-1}$, $\Omega_\Lambda = 0.73$,
$\rm \Omega_M = 0.27$ and Vega magnitudes through out.

\section{Data}

In order to investigate the general nature of flaring phenomena in galaxies it is 
necessary to begin with a large, clean, sample of sources. Detection sensitivity 
requires that the sources have been monitored with the timespan and cadence required 
to discern such events from other types of variability and noise sources. The 
combination of photometry from the Catalina Surveys with SDSS spectroscopy 
provides us with this for more than a million galaxies.

\subsection{Catalina Photometry}

The Catalina Surveys consist of two separate surveys that use the 
same data for different purposes. These consist of the Catalina Sky 
Survey\footnote{http://www.lpl.arizona.edu/css/} that focuses on the 
discovery of Near-Earth Objects (NEOs, Larson et al.~2003), and 
the Catalina Real-time Transient Survey (CRTS\footnote{http://crts.caltech.edu/})
which searches for optical transients (Drake et al.~2009).

The NEO survey began taking observations with three telescopes in 2003.
Each telescope was run as a separate sub-survey. These consisted of the 
0.7m Catalina Schmidt Survey (CSS\footnote{Note: CSS in this work only 
relates to this telescope}) and the 1.5m Mount Lemmon Survey (MLS), both 
in Tucson, Arizona, and the 0.5m Siding Spring Survey (SSS) at Siding Spring, 
Australia. CRTS began by processing data from CSS in 2007 and later 
added data from MLS and SSS.

Each of the Catalina telescopes takes observations using a grid of fields 
that tile the sky. These surveys, when combined, cover most of the sky between 
declinations $\delta = -75$ and +65 degrees. However, regions within
$\sim$15$\arcdeg$ of the Galactic plane are generally avoided due to crowding.

The original MLS and CSS $\rm 4k^2$ cameras have recently been upgraded 
to increase their field-of-view. However, here we only consider data taken 
in the original configuration, where MLS and CSS survey images covered 
1.2 $\rm deg^2$ and 8.2 $\rm deg^2$, respectively. We do not 
consider SSS data since this generally does not overlap with the
SDSS footprint.

Almost all Catalina observations are taken in sequences of four 
images covering the same field each 10 minutes. 
Isophotal photometry is obtained for all sources in each observation 
using the SExtractor photometry program parameter "MAG\_ISOCOR" (Bertin \& Arnouts 1996).
All of the Catalina images are taken without a filter and are calibrated 
to a pseudo-$V$ magnitude ($V_{\rm CSS}$) using a few dozen pre-selected 
standard stars in each field. Further details of the photometric 
calibration and transformations to standard systems are given in 
Drake et al.~(2013).

Within the Catalina photometric dataset we concentrate on photometry
taken between 2007 January 1 and 2016 April 28. We do not consider earlier 
Catalina photometry since some of it was measured using a larger aperture 
that caused inaccurate photometry when neighbouring sources were present.

Most of the photometry used in this work is publicly available as part of 
Catalina Surveys Data Release 2 (CSDR2)\footnote{http://catalinadata.org}.
However, some additional more recent photometry is also included, which 
will soon be made available as part of CSDR4 (Drake et al., in prep.).

\subsection{SDSS Spectra}

The SDSS has operated for well over a decade and undertaken multiple
spectroscopic surveys.  For this analysis we selected spectroscopic data
from SDSS DR13 (Albareti et al.~2017) as our reference set. This was the
most recent SDSS data release at the time of the analysis and contains 
4.4 million optical spectra of QSOs, galaxies, and stars.  This dataset
includes spectra from the SDSS legacy survey (York et al.~2000), as well as
those from the new spectrograph used for the Baryon Oscillation
Spectroscopic Survey (BOSS) (Dawson et al.~2016), and the Sloan Extended
Quasar, ELG, and LRG Survey (SEQUELS) program (Albareti et al.~2017 and
references therein).

The original SDSS legacy survey spectra have a wavelength range of 380-920 nm 
with resolution $R\sim2000$ and $3\arcsec$ fibres, while the BOSS spectra 
cover 360-1000 nm with $R\sim2000$ and $2\arcsec$ fibres.
The legacy survey targeted what is known as the Main Galaxy Sample (MGS).
This consists of bright galaxies with Petrosian magnitude $r_P \le 17.77$ 
(Strauss et al.~2002), a Luminous Red Galaxies (LRG) sample out to $z=0.5$,
and a QSO sample (Schneider et al.~2010). The more recent BOSS survey targeted 
fainter, more distant galaxies ($0.15 < z < 0.7$) and QSOs ($z > 2.15$) than 
the SDSS Legacy survey (Dawson et al.~2016).

\subsection{Galaxy Sample Selection}

Among the full sample of SDSS spectra available in DR13, 2.6 million sources 
are classified as galaxies. A small number of the SDSS
spectra are repeat observations of the same source.  Of the SDSS galaxies, 
we select the 1.7 million sources with redshifts $z < 0.5$.  We
specifically limit our sample to the nearer sources since we are insensitive
to most outbursts in more distant galaxies.  For example, a magnitude 
$V_{\rm CSS} = 19.5$ galaxy at $z=0.5$ has $M_V =-22.5$
(ignoring extinction). Thus, to detect an outburst in such a galaxy, the
event would need to be among the few brightest known superluminous
supernovae. Additionally, since more distant galaxies are, on average,
fainter, they tend to have lower signal-to-noise ratio spectra.

The detection of flaring events in the presence of AGN has already been
addressed for CSS data for 900,000 source by Graham et al.~(2017). That work
revealed many AGN undergoing large flaring events. Here we aim to
test whether similar outbursts also occur in quiescent galaxies. However,
we do not remove the small fraction ($< 4\%$) of SDSS galaxies that are 
also spectroscopically classified as containing AGN.

Matching the locations of the $z < 0.5$ SDSS DR13 spectroscopic galaxy
sample to those among the 200 million sources in the CSS catalog, 
we find $1.46$ million matches within $5\arcsec$. Since our aim is to find 
outbursts in the lightcurves of these sources, we sub-select the galaxies 
with 20 or more nights of CSS photometric observations.
This selection reduces the total number of SDSS galaxies with CSS lightcurves to 
1.43 million. Some of the matched galaxies are duplicates due to overlap 
between CSS fields. In the final dataset there are $1.42$ million unique 
CSS sources having SDSS spectra.

We also matched the nearby ($z < 0.5$) SDSS galaxy sample to sources from
the MLS survey. This data only covers objects near the ecliptic, where there 
are only 377,000 matches with SDSS. The MLS photometry is also much 
sparser than CSS. Among these matches only 152,000 have more 
than 20 nights of MLS observations. Since the number of SDSS galaxies 
with MLS lightcurves is only $\sim10$\% that of the CSS set, we 
will primarily concentrate our analysis on the CSS sample.

\section{Candidate Selection}

In order to select a set of photometric outbursts, among the sample
of 1.4 million spectroscopically confirmed galaxies with significant 
CSS photometry, we must first understand the range of photometric uncertainty 
and the extent of individual object variability. Once we understand these, 
we can then identify and remove sources of noise to minimize any impact 
they might have on the results.

\subsection{Determining Photometric Uncertainties}

\begin{figure*}{
\includegraphics[width=84mm]{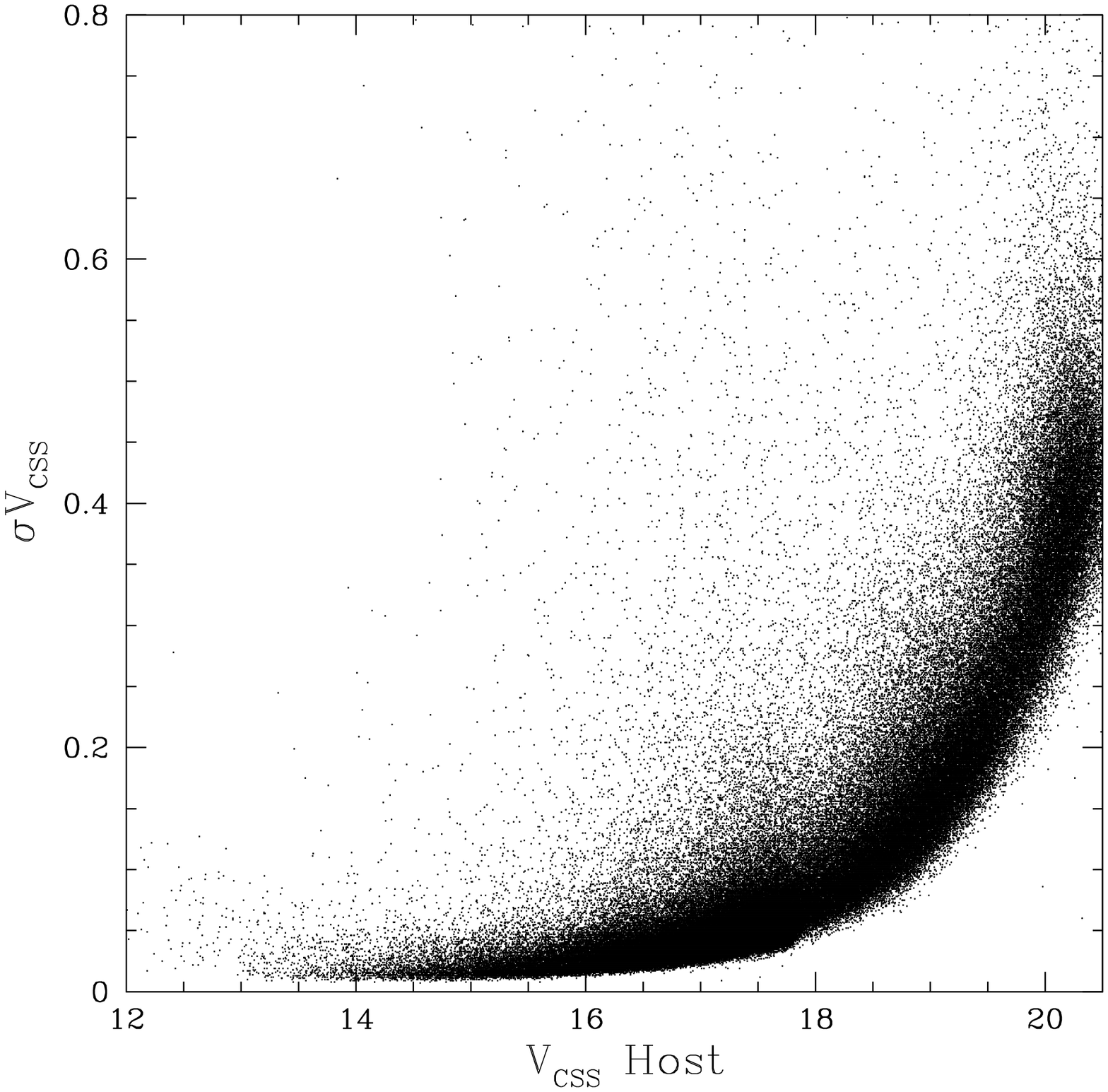}
\includegraphics[width=84mm]{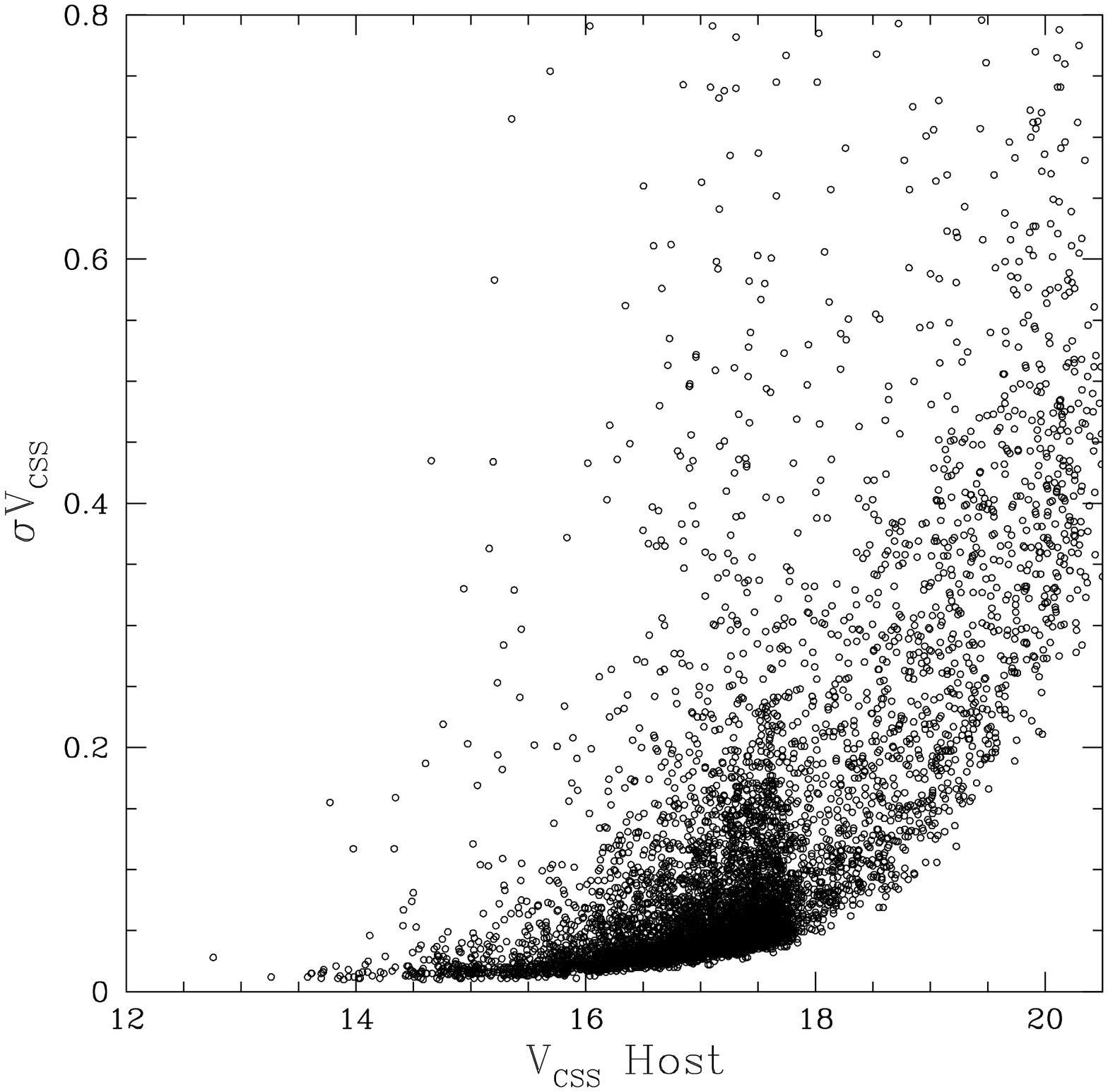}
\caption{\label{GalS}
CSS photometric variability.
In the left panel, we plot the scatter for a sample of
200,000 spectroscopically confirmed SDSS galaxies
with $z < 0.5$.
In the right panel, we show the scatter for galaxies that 
are near bright stars or are close to the locations where 
we expect to see bleed trails from very bright stars.
}
}
\end{figure*}

Our main goal in this work is to find galaxies exhibiting flaring events
among the SDSS sample of spectroscopically confirmed galaxies. Such events
may be due to supernovae, tidal disruption events (Rees 1988), AGN activity,
microlensing (Lawrence et al.~2016), or other hitherto unknown sources.
Since our aim was to find outbursts in general, we decided to minimize 
assumptions about the shapes of the events. Thus, our initial selection 
of candidates was based on statistical significance rather than model 
fitting.

In order to select galaxies exhibiting significant changes in brightness 
it is necessary to understand the underlying degree of scatter in the
photometric data. Based on reduced $\chi^2$ values for fits to periodic 
variable lightcurves, Drake et al.~(2014) showed that the photometric 
uncertainties given for CSS photometry in CSDR2 are generally overestimated 
for bright sources and underestimated for fainter sources.  
Graham et al.~(2017) further quantified the values required to correct 
the average magnitude-dependent effect. Nevertheless, since these 
corrections are averaged over $\sim200$ million objects distributed 
across most of the sky, they fail to fully represent the low level 
variations in individual lightcurves due to differences in crowding, 
airmass, weather, etc.
Thus in this work, rather than correcting the uncertainties using the
values in Graham et al.~(2017), we instead measure the scatter in the 
photometry for each lightcurve as an initial estimate of the true uncertainty.

As a starting point, we assume the uncertainties are normally distributed
and use the standard deviation for each source as the measure of scatter.
In the left panel of Figure \ref{GalS}, we plot the distribution of CSS
photometric uncertainties for a sub-sample of 200,000 of the SDSS
galaxies. Evidently, galaxies brighter than $V_{\rm CSS} = 16$ typically 
have $\sigma < 0.04$ mags and at $V_{\rm CSS}\sim 20$ galaxies have 
$\sigma\sim0.5$ mags.

We note that the presence of real underlying variability, such as due to AGN, 
will tend to skew or broaden the observed distributions as will
outliers due to noise sources. To mitigate the effect of outliers
we perform a $3\sigma$ clipping on the measurements.  We then
recalculate the scatter from the clipped distribution. The clipping process
typically only removes a few outliers from each lightcurve.  The outliers
themselves are mainly bright points caused by artifacts, such as saturated
columns, internal reflections and satellite trails, as discussed in the next
section.

We do not expect that long-timescale variability will be significantly 
affected by the clipping process. For example, the long-timescale
variability typical of AGN (Macleod et al.~2012) will tend to broaden 
the distribution beyond the true photometric uncertainties, but few if 
any points will be removed.

\subsubsection{Sources of Photometric Outliers}

As noted above, our simple statistical selection of outbursts is only
dependent on the distribution of photometric measurements.  The lightcurves
can be affected by both real outbursts as well as noise sources within the
images. These two sources both generally produce a long positive tail in the
measurement distributions. It is therefore necessary to separate the noise
sources from the actual outbursts.

One of the main sources of photometric outliers are image artifacts due 
to very bright stars.  Such stars can cause saturated lines, blooms and halos 
that can affect the measurements of both nearby and distant sources.
Additionally, the effects caused by bright stars can occur intermittently,
since a bright star may not saturate an image taken in poor seeing, 
or when the sky background level is low.

As part of the CRTS transient search processing, all CSS data is 
run through a pipeline that automatically masks objects in close 
proximity to very bright stars (Drake et al.~2009). 
The values used to veto transient candidates near bright
stars are the same as those used to cull AGN flare candidates
near bright stars as presented in Graham et al.~(2017). That 
process was also adopted to veto flare candidates in this work.
In the right panel of Figure \ref{GalS}, we show the scatter for the 
small fraction of galaxies that are in close proximity to saturated stars. 
This small group ($<1\%$) of galaxies contributed a large fraction 
of all sources having significant scatter.
In addition, the brightest stars also introduce noise into the lightcurves
of much more distant sources via line-bleeds.  Indeed, since the
field-of-view of the Catalina CSS camera spans more than two degrees, these
line-bleeds also can affect lightcurves of sources that are very far from
the saturated stars.

Inspecting a number of the full Catalina images we found that the 
line-bleeds are generally contained to within tens of arcseconds 
along right ascension (even for stars with $V_{\rm CSS} < 8$). 
A smaller effect was also measured along declination for saturated stars. 
By using the known locations of bright stars in the Tycho catalog we 
compiled a list of galaxies within our sample that could be affected 
by bleeds from saturated stars lying anywhere within their field.
This list contains only $\sim$17,000 galaxies among the 1.4 million 
CSS-SDSS galaxy sample.

Considering the full observational situation it is self-evident that this
list would not contain all affected galaxies. For example, the degree of
saturation-induced bleeding alone depends on the colour of the star as well
as changes in the background sky level, etc.  In fact, the complexity and
diversity of the effects caused by bright stars limits our ability to
algorithmically mitigate the sources of all outliers. Inspection of the
lightcurves and images showed us that many additional lightcurves were
affected by related effects as well as unrelated noise sources.
Specifically, there were many cases where galaxies were not in close
proximity to a very bright star, but rather multiple moderately bright
sources. There were also lightcurves with photometry that was affected by
reflections, cosmic rays and satellite trails.  Thus, rather than simply
removing galaxies that were potentially affected by neighbour saturated
stars, we concluded that it would eventually be necessary to inspect the
images for every single outburst candidate.

The need for the final by-eye image verification was not surprising, since
even modern transient surveys employing machine learning classification
techniques still require human {\it scanners} as the final step in transient
confirmation. This is due to the fact that both transients and artifacts are
generally rare (sometimes unique). Their scarcity often places them in the
same poorly defined region of the detection parameter space, where the
measured observables are insufficient for a unique classification without
the addition of contextual data (such as can be gained from the images,
historical lightcurves, multi-wavelength data, etc.).

\subsection{Initial candidate selection}

\subsubsection{Information from Known Events}

In order to investigate the methods and thresholds that might be appropriate
for finding outbursts, we decided to extract the lightcurves for every
galaxy with an SDSS spectrum and a known supernova candidate.  Although we
are not specifically searching for supernovae, such sources clearly provide
the largest and most well understood population of outburst events within
galaxies.

We matched the online Rochester SN catalog (Gal-Yam et al.~2013),
consisting of $\sim$25,000 known SN candidates, with the spectroscopically
confirmed SDSS galaxies in our sample. In total, this dataset consisted of
3,131 unique SDSS galaxies with SN candidates and CSS lightcurves.  Due to
repeated spectroscopic observations, there are actually 3,464 individual
SDSS spectra for these hosts. Additionally, due to overlap between CSS
fields, these galaxies match 3,207 separate CSS lightcurves.

Of the 3,131 supernova candidates, a large fraction are
undetectable within the Catalina lightcurves as they occurred 
before the galaxy was monitored, or during observing gaps. 
Overall, only 1,732 of the SN candidates in SDSS confirmed 
galaxies were discovered within the range of dates analyzed here.

For each of these galaxies we extracted the CSS lightcurve. We then plotted
and inspected each. Many of the SN candidates show no sign within CSS
lightcurves since they are well offset from the cores of their hosts.
Nevertheless, even when there is no evidence for the supernova candidates in
the CSS lightcurves, these galaxies can still provide a useful sample for
testing and understanding the properties of SN host galaxies.  The set of
hosts where the SN candidates were clearly present within the CSS
lightcurves provided us with a means for determining whether our detection
method and threshold was actually selecting the events.

\subsubsection{Filtering candidates}

Overall, based on our sigma clipping we know that outliers occur at a rate 
of around 1\% of the data. Nevertheless, in some cases there can be dozens of outliers.
Thus, with more than a million lightcurves, the selection of outburst candidates 
based on a single night is impractical. Significant additional evidence would 
be required to believe the validity of any such candidates.
Likewise, the expected false positive rate due to two consecutive
nights of high measurements is also too high to have confidence
in candidates. 
However, the probability of a third night containing an outlier
in a sequence of between 20 and 200 nights is highly unlikely 
($\lsim 0.1\%$, assuming their occurrence is uncorrelated). Thus, our 
initial criterion for selecting outburst candidates was to require that each 
candidate was detected on at least three contiguous nights of observations.

This selection enables us to mitigate most of the problems due to random
artifact occurrences. Nevertheless, in some cases outliers due to spurious 
sources are also correlated in time~--for example, due to lunar cycles, 
seasonal variations in observational airmass, weather patterns, etc.  
Thus, rather than, for example, simply selecting sources that are 
outliers based on the presence of, for example, three $>3\sigma$ contiguous 
detections, we decided to estimate the likelihood of each detection 
based on a normal distribution. 
To simplify the process we only calculate the probabilities for sequences 
of measurements where there are at least three values $>1\sigma$ above 
the median brightness. Such sequences occur in a random sequence of normally 
distributed data $\sim0.4\%$ of the time.

\subsubsection{Modeling Coincidences}

\begin{figure*}{
\includegraphics[width=84mm]{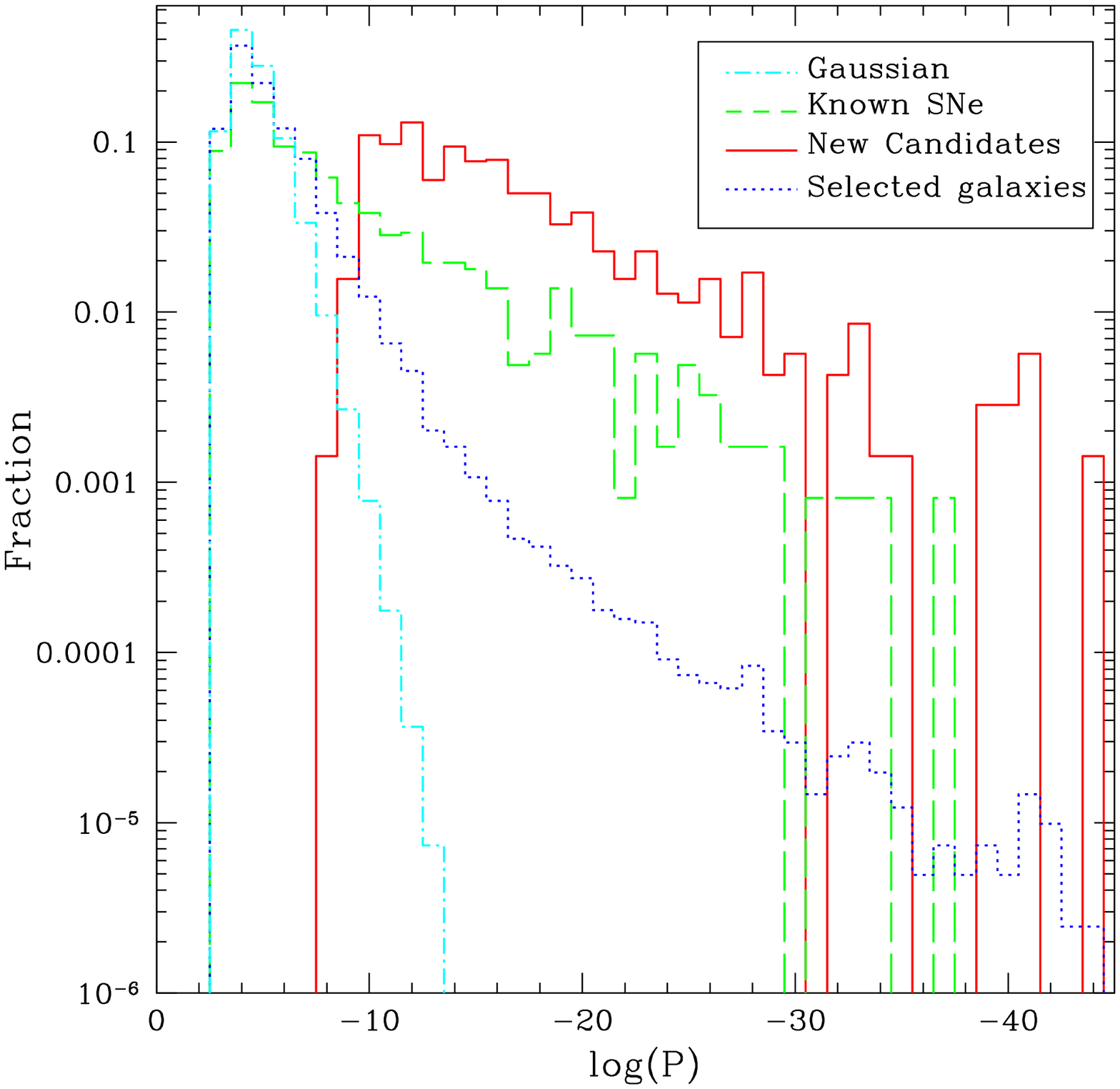}
\includegraphics[width=84mm]{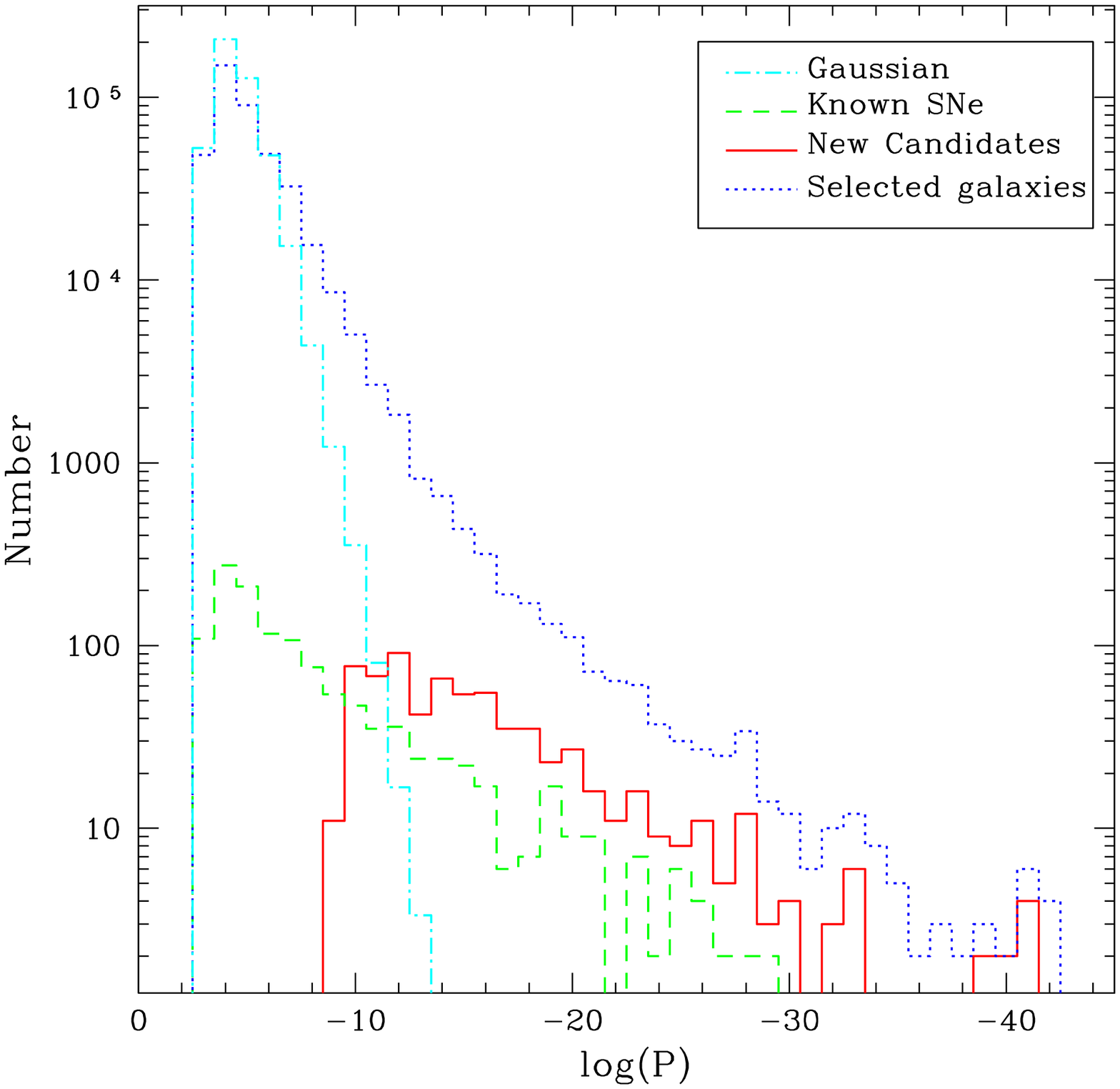}
\caption{\label{Signif}
The distribution of candidate outburst significance.
The dotted-blue lines show the distribution of detection 
significance values for the 400,000 event candidates with three
contiguous detections above $1\sigma$. 
The cyan dash-dotted line gives the number expected 
based on a purely Gaussian distribution of measurements.
The green histogram
shows the distribution for the 1,232 galaxies with
previously known SN candidates and three contiguous 
detections above $1\sigma$. The solid-red histogram
shows the distribution of significance for the newly
discovered outbursts selected with $log(P) < -8.5$.
The left panel shows the fractions of the total,
while the right shows the actual numbers of 
sources.
}
}
\end{figure*}

We carried out simple numerical simulations involving the production
of a few hundred thousand lightcurves with purely Gaussian deviates.
We found that $\sim 30\%$ of these lightcurves meet our initial 
selection criteria.
Such a low threshold enables us to potentially retain some sensitivity 
to selecting events with long timescales when the individual nights 
have only marginal significance. However, it was obvious that much 
more stringent criteria were required to remove the coincidental 
detections.

The initial selection process was run on the known SN candidates 
as well as on the full set of 1.4 million galaxy lightcurves. 
In total, $\sim$400,000 of the lightcurves had at least three contiguous 
$1\sigma$ deviations (in good agreement with our expectations). 
For all the lightcurves meeting the initial threshold, we then estimated 
the probabilities, $P$, of each set of high points randomly occurring. In 
each case the combined nightly average magnitudes were assumed to be 
normally distributed. 

In Figure \ref{Signif}, we plot the distribution of expected probabilities
for each of the candidate outbursts in the initial selection as well as those
for the 1,232 previously known SNe candidates that meet our initial
selection criteria. This plot also shows the number of objects and probabilities
expect based on our simple numerically simulated lightcurves with pure normally
distributed data. One can see that for events with $\rm log(P) > -11$, 
the number of candidates selected is consistent with that expected
from our simulated normally distributed data. However, the actual
distribution of probabilities for the initial 400,000 selected sources is very
different from the pure Gaussian simulation. The observed distribution shows
a strong tail. This is due to a combination of real variability, the
presence of outliers, and the heteroscedastic nature of photometric
uncertainties.

\subsection{Final Candidate Selection}

As noted above, based on our initial investigation there was a clear need 
to inspect the data relating to each candidate. To decrease the number
of spurious outbursts, we decided to inspect both the lightcurves 
and the images of each candidate.

\subsubsection{Lightcurve Inspection}

To select good candidates from the initial large set, we firstly 
inspected the lightcurves of a few hundred random galaxies as well 
as all 3,207 galaxies containing previously known SN candidates. 
We payed particular attention to the lightcurves of galaxies 
where the SN candidate were know to occur during our analysis 
period.

This initial inspection led us to select only candidates where the 
chance of random occurrence was $\rm log(P) < -8.5$. This threshold 
was determined based on requiring that known SN candidates should be 
selected via the presence of the actual outburst.
We determined that events with peaks occurring within 100 days 
of the candidate's discovery date were likely due to that event.
In such cases there is only a $\sim$6\% chance of any of these events 
occurring randomly given that span of the observations is $\sim$3,500 days. 
We found that most of the known SN candidates that were among 
the initial 400,000 objects, though a number were missed by the $\rm log(P) < -8.5$ 
selection as they did not exhibit significant sign of the SN in the lightcurve.

Having decided on our probability threshold, we inspected the lightcurves of 
every source with $\rm log(P) < -8.5$.  Of these, $> 90\%$ of the candidates 
were rejected simply by considering the presence of variability (due to real
sources or noise artifacts) in the lightcurves.  We selected a refined set
of candidates by further considering the shapes of each outburst candidate,
additional measurements surrounding each candidate event and event 
coincidence in MLS data. Overall, this process led to an initial set 
of 1,090 new outburst candidates.

\subsubsection{Image Inspection}

\begin{figure}{
\includegraphics[width=84mm]{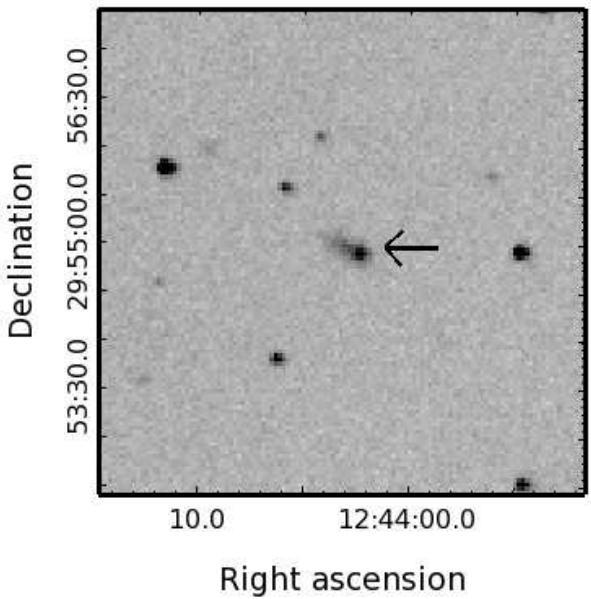}
\caption{\label{SNexamp}
One of the four, UT 2012 January 9 CSS images showing the 
presence of a new source offset from the core of the galaxy 
KUG 1241+301. This object is responsible for outburst 
candidate CRTS\_OBC\_341.
}
}
\end{figure}

After our initial selection of candidates based on lightcurve data, 
we inspected the images associated with the peak for each candidate. 
This process enabled us to remove candidates due to image artifacts
(such as saturated columns and blending).

For every candidate we extracted cutouts $5\arcmin$ across for the 
images taken within 10 days of the photometric peak. 
For the initial set of 1,090 CSS outburst candidates this consisted 
of 7,638 separate images. After removing events due to image artifacts 
we were left with a final set of 692 outburst candidates.

While examining the images of the outburst candidates we noted that a small 
fraction ($\sim10$\%) clearly showed the presence of a new source that was 
separated from the core of the host galaxy.
We were able to verify that these were indeed real objects by comparing 
the outburst images with the deeper, higher resolution, images of the hosts
from SDSS. We note that all of the SDSS images were taken prior to the 
CSS data, so the presence of new sources gives us very high confidence that 
these detections are genuine. Furthermore, since TDEs and AGN flares are 
only expected in the cores of galaxies, these particular sources are all 
likely to be supernovae.
In Figure \ref{SNexamp}, we show an example of likely supernova 
CRTS\_OBC\_341. Here the new source is clearly seen offset from 
the core of the galaxy KUG 1241+301.

For the bulk of the outburst candidates, the lack of a new 
source is expected since sources that are resolved from their 
host galaxies should be detected as separate new sources.
It may be possible to resolve additional outbursts from their
hosts using image subtraction. However, since most of the host 
galaxies with new candidates appear relatively compact, and the 
CSS pixels are large ($2.5\arcsec$), this is unlikely to lead 
to many with statistically significant offsets.

\section{Outburst Candidates}

\begin{table*}
\caption{Properties of CSS outbursts\label{cssout}}
\begin{minipage}{186mm}
\begin{tabular}{@{}crrccrrrrrc}
\hline
ID & RA (J2000, deg) & Dec (deg) & $V_{\rm CSS}$ & $M_V$ & $\rm MJD_{peak}$ & length (d) & Signif & log(P) & Nights & Quality\\
\hline
CRTS\_OBC\_1 &   139.3122 &     4.1337 & 18.01 & -15.55 & 54038.5 &  174 & 70.91 & -30.91 & 10 & I  \\
CRTS\_OBC\_2 &   163.2952 &    27.7206 & 19.02 & -16.53 & 54062.5 &  272 & 33.41 & -18.69 & 8 & I  \\
CRTS\_OBC\_3 &   165.9949 &    26.3342 & 18.37 & -20.37 & 54062.5 &  405 & 49.36 & -32.95 & 12 & I$^b$\\
CRTS\_OBC\_4 &    24.7624 &     6.4963 & 18.78 & -18.98 & 54108.1 &  246 & 14.08 & -13.05 & 4 & II  \\
CRTS\_OBC\_5 &   131.8185 &    31.8859 & 18.99 & -20.04 & 54117.3 &  235 & 44.67 & -39.27 & 13 & I  \\
CRTS\_OBC\_6 &   175.2558 &     5.2601 & 17.95 & -17.73 & 54124.5 &  102 & 24.11 & -15.78 & 5 & I$^a$ \\
CRTS\_OBC\_7 &   146.3246 &    34.0959 & 18.85 & -17.48 & 54127.2 &   82 & 17.27 & -10.22 & 4 & III  \\
CRTS\_OBC\_8 &   155.7226 &    33.3855 & 19.25 & -17.20 & 54127.2 &  113 & 13.57 & -12.31 & 6 & I$^b$\\
CRTS\_OBC\_9 &   213.9948 &     2.7104 & 17.70 & -19.94 & 54128.5 &  140 & 31.70 & -24.89 & 7 & I  \\
CRTS\_OBC\_10 &   172.4056 &    14.4719 & 18.99 & -18.44 & 54139.4 &   87 & 29.32 & -15.63 & 6 & I$^a$ \\
\hline
\end{tabular}
\end{minipage}\\
\begin{flushleft}
{\it The full table will be available online.}\\
Column 1: CRTS outburst candidate ID.\\
Columns 2 and 3: Source coordinates.\\
Columns 4 and 5: Apparent and absolute magnitudes
for the peak of the outburst, respectively.\\
Columns 6: MJD of the outburst peak.\\
Column 7: Timespan over which the candidate outburst 
was detected about $1\sigma$.\\
Column 8: Total significance in sigma of the detections
during the outburst timespan.\\
Column 9: Probability of false detection assuming normally
distributed data.\\
Column 10: Number of nights when the outburst was detected
above $1\sigma$.\\
Column 11: Quality of the outburst candidate based the inspection 
of lightcurves and images as well as the presence of detections 
in MLS data. The values are as follows: $\rm{I}$ --- high confidence 
event; $\rm{II}$ --- moderate confidence 
event; $\rm{III}$ ---- low confidence event.
\\
$^a$ outburst is also detected in MLS observations.
\\ 
$^b$ source is resolved from the host galaxy.
\\ 
$^c$ long timescale event.\\ 
\end{flushleft}
\end{table*}

In Table \ref{cssout}, we present the parameters for each of the 692
new CSS outburst candidates. This table includes the length and 
significance of each outburst candidate as well as an assessment
of the quality of each candidate based on the inspection of
the lightcurve and images. Most of the 80 candidates with 
a quality value of $\rm III$ have $log(P) > -12$ 
and are thus potentially statistical coincidences. In some
cases a candidate is deemed of higher quality than the
probability would suggest because it is also seen in
MLS photometry.

\begin{figure*}{
\includegraphics[width=84mm]{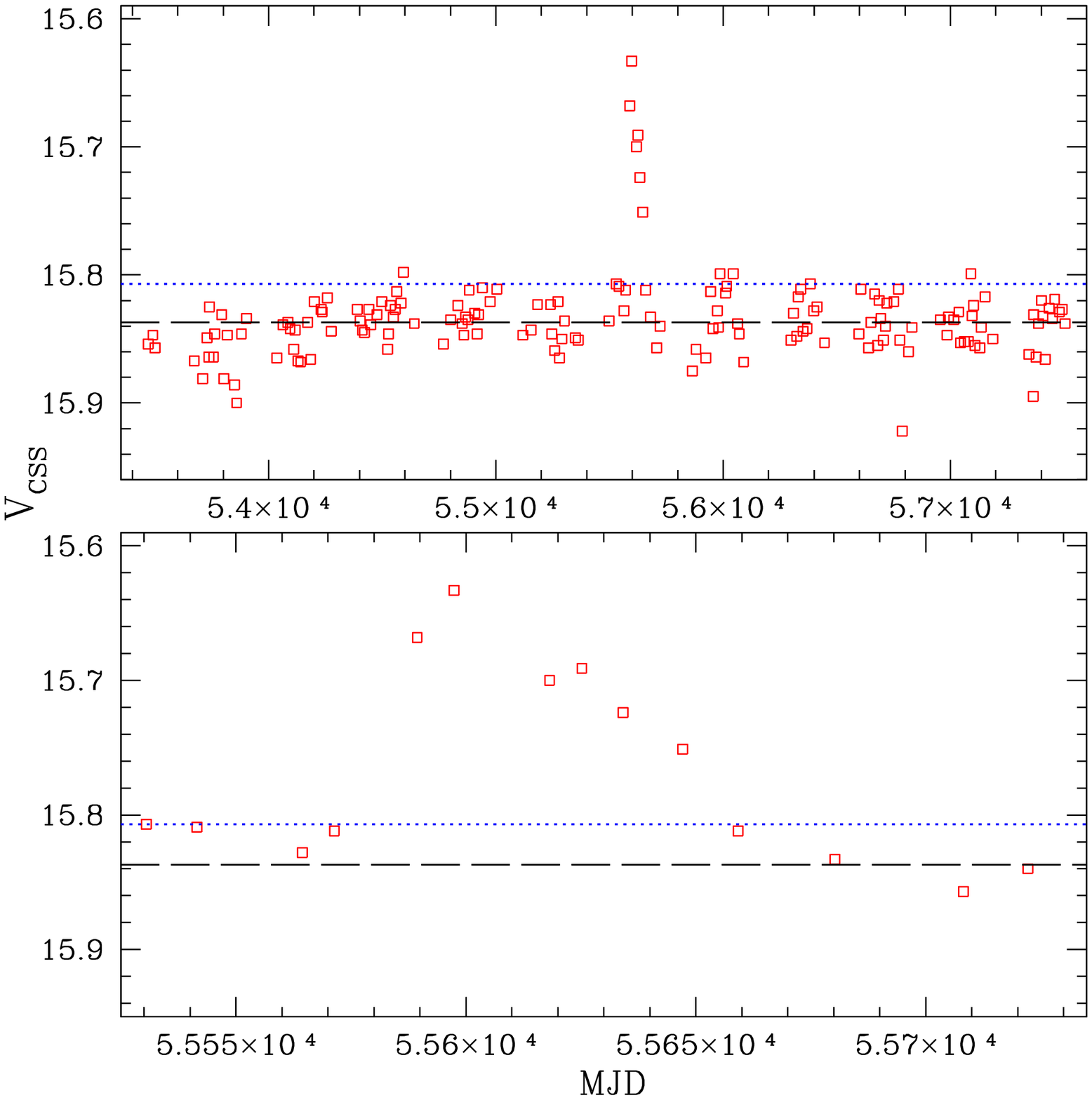}
\includegraphics[width=84mm]{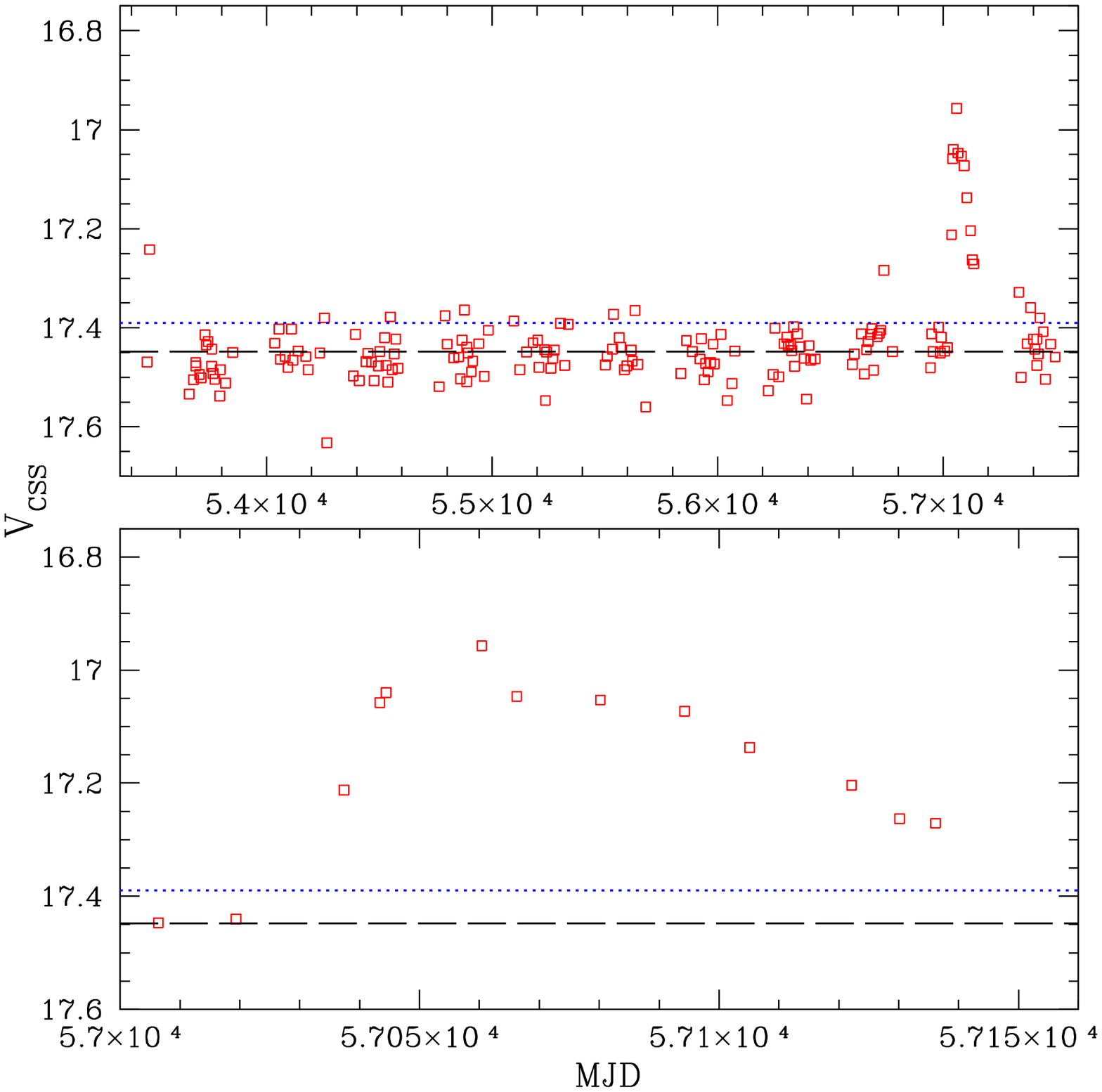}
\caption{\label{CRTS_OBC_284}
Outburst candidate lightcurves.
In the left panels, we show the nightly-averaged CSS lightcurve of 
CRTS\_OBC\_284 and in the right panel, we show that of CRTS\_OBC\_602. 
The host galaxies are SDSS J104018.90+230525.5 ($z=0.046$) and 
SDSS J091620.42+284428.0 ($z=0.064$), respectively.  
The upper panels show the full lightcurve and the 
lower panels show the span of each outburst.
The blue dotted-line shows the $1\sigma$ threshold.
The long dashed-line gives the median magnitude.
}
}
\end{figure*}

\begin{figure*}{
\includegraphics[width=84mm]{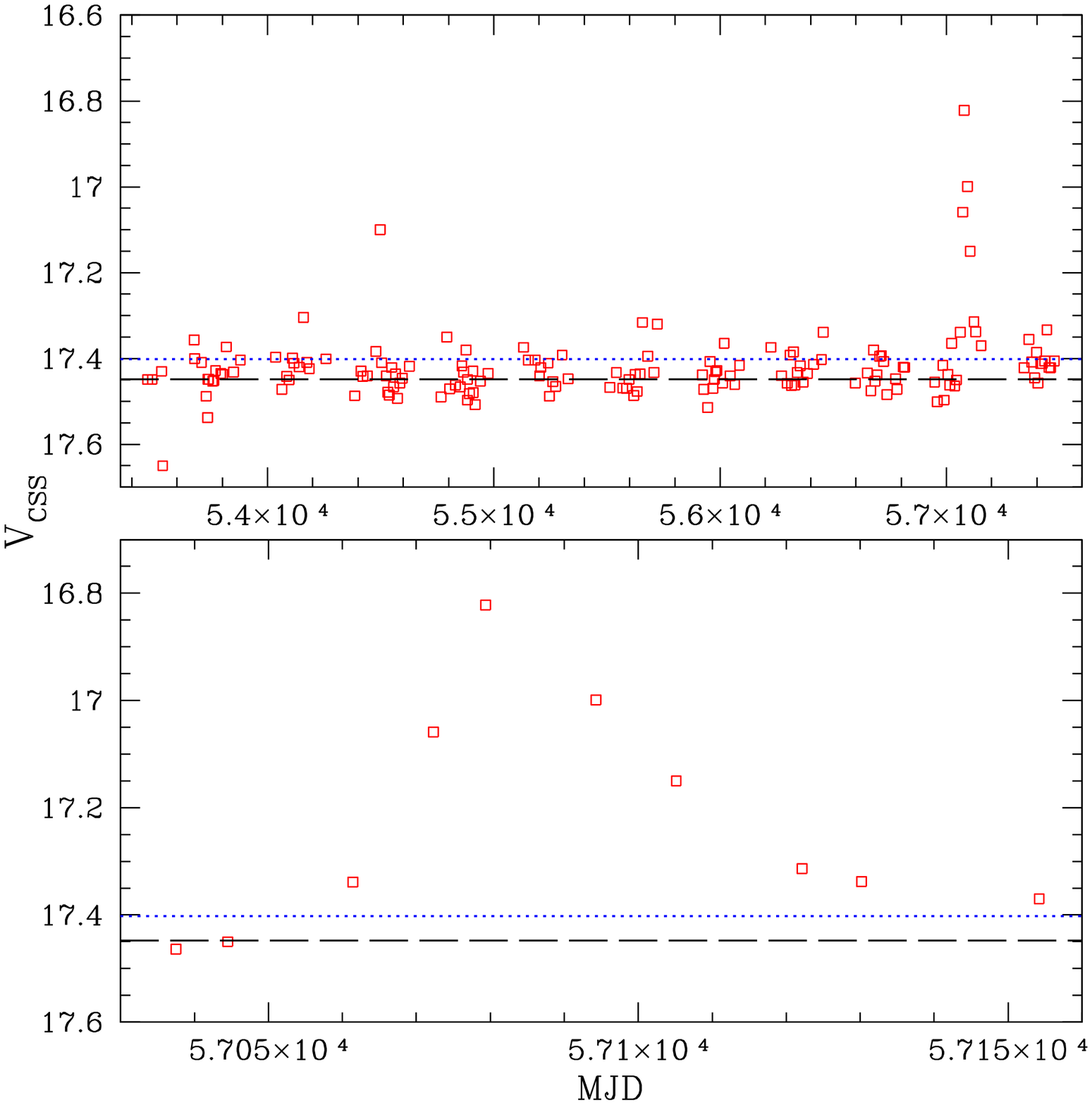}
\includegraphics[width=84mm]{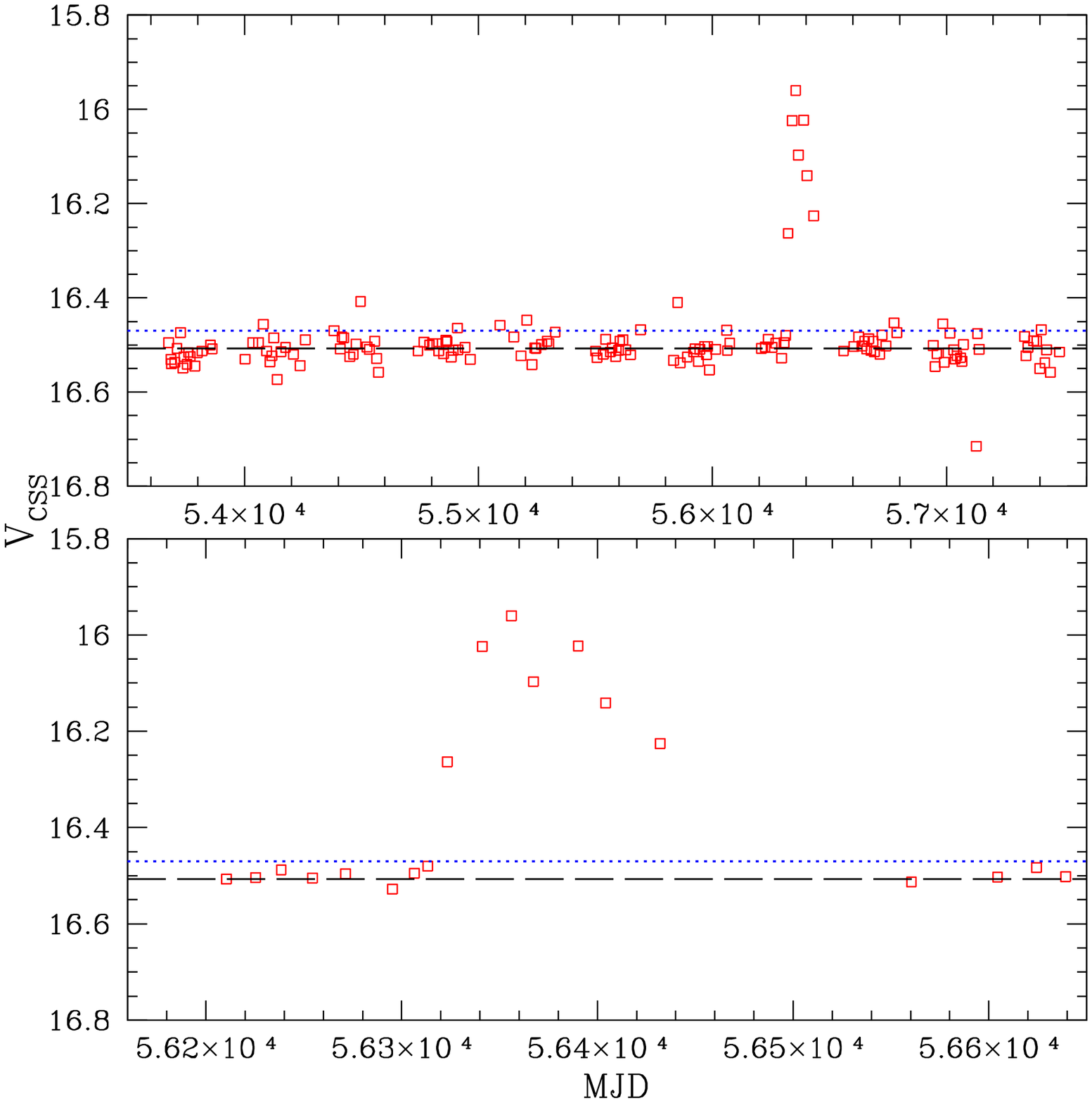}
\contcaption{
Nightly-averaged lightcurves for CRTS\_OBC\_613 (left) 
and CRTS\_OBC\_451 (right).
The host galaxies are SDSS J102352.98+292333.4 ($z=0.049$)
and SDSS J084904.74+342045.3 ($z=0.059$), respectively. 
}
}
\end{figure*}

\begin{figure*}{
\includegraphics[width=84mm]{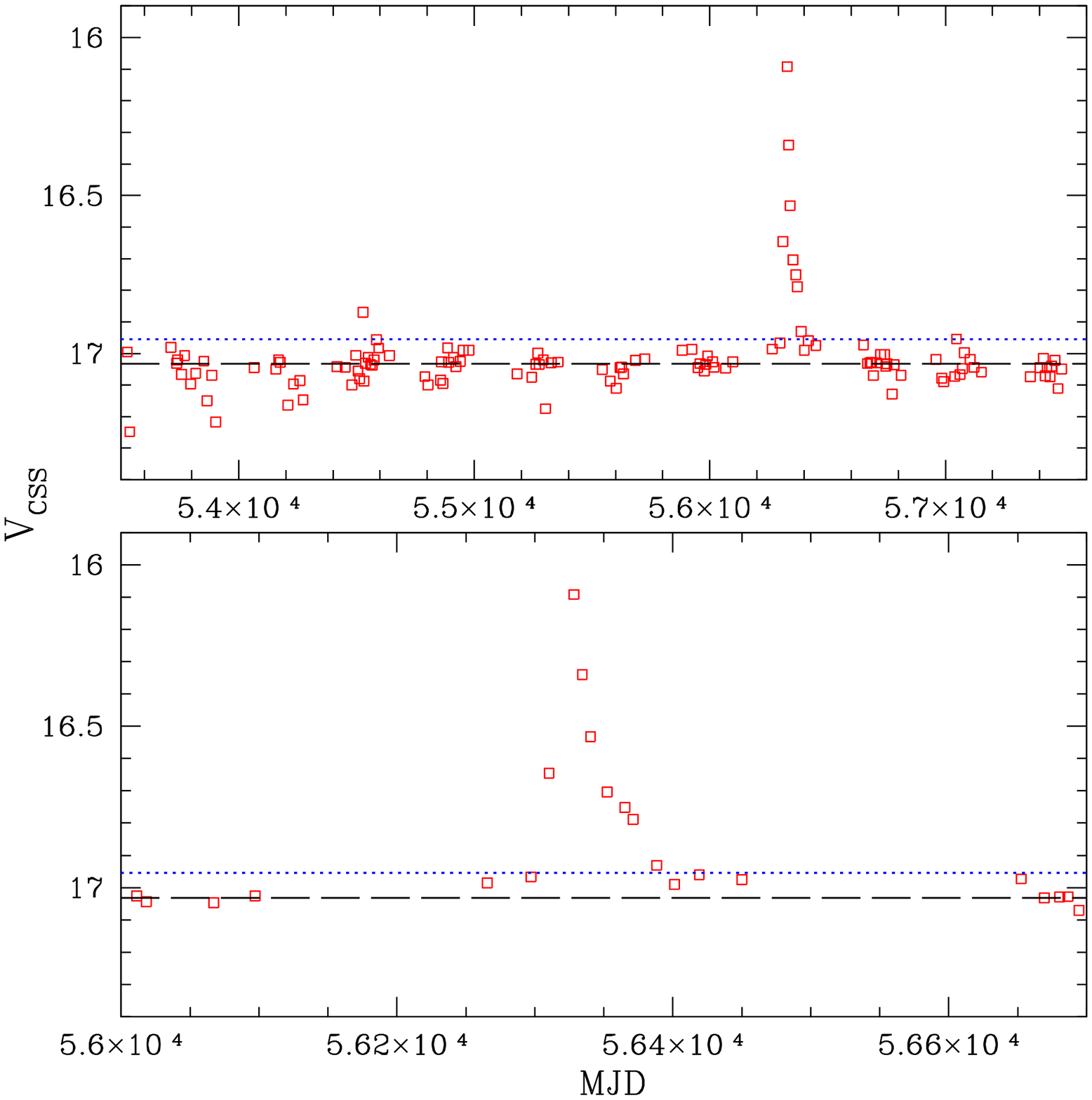}
\includegraphics[width=84mm]{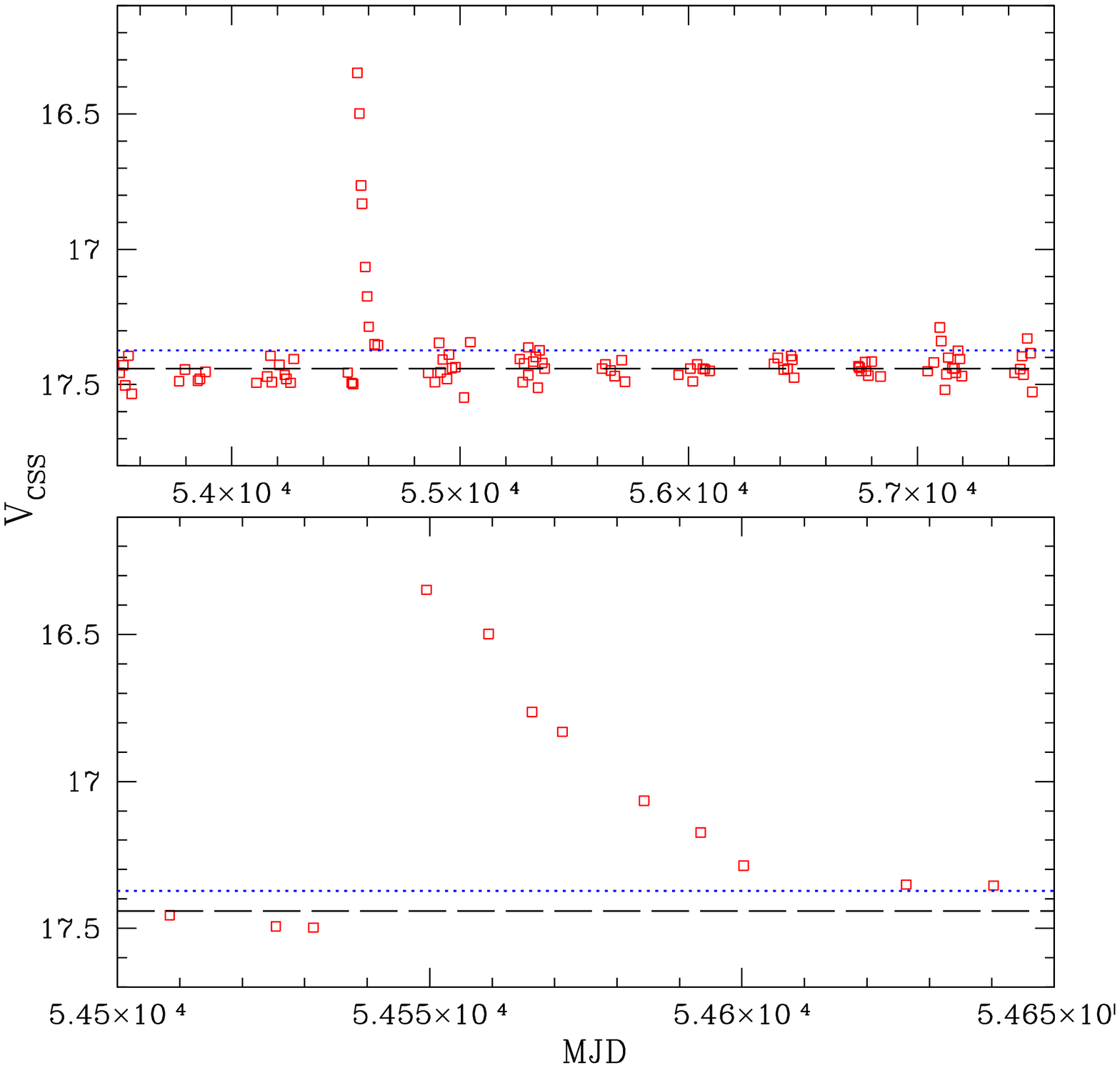}
\contcaption{
Nightly-averaged lightcurves for CRTS\_OBC\_436 (left) 
and CRTS\_OBC\_102 (right). The host galaxies
are SDSS J111452.35+362516.6 ($z=0.024$) and
SDSS J143345.61+344831.9 ($z=0.035$), respectively.
}
}
\end{figure*}

\begin{figure*}{
\includegraphics[width=84mm]{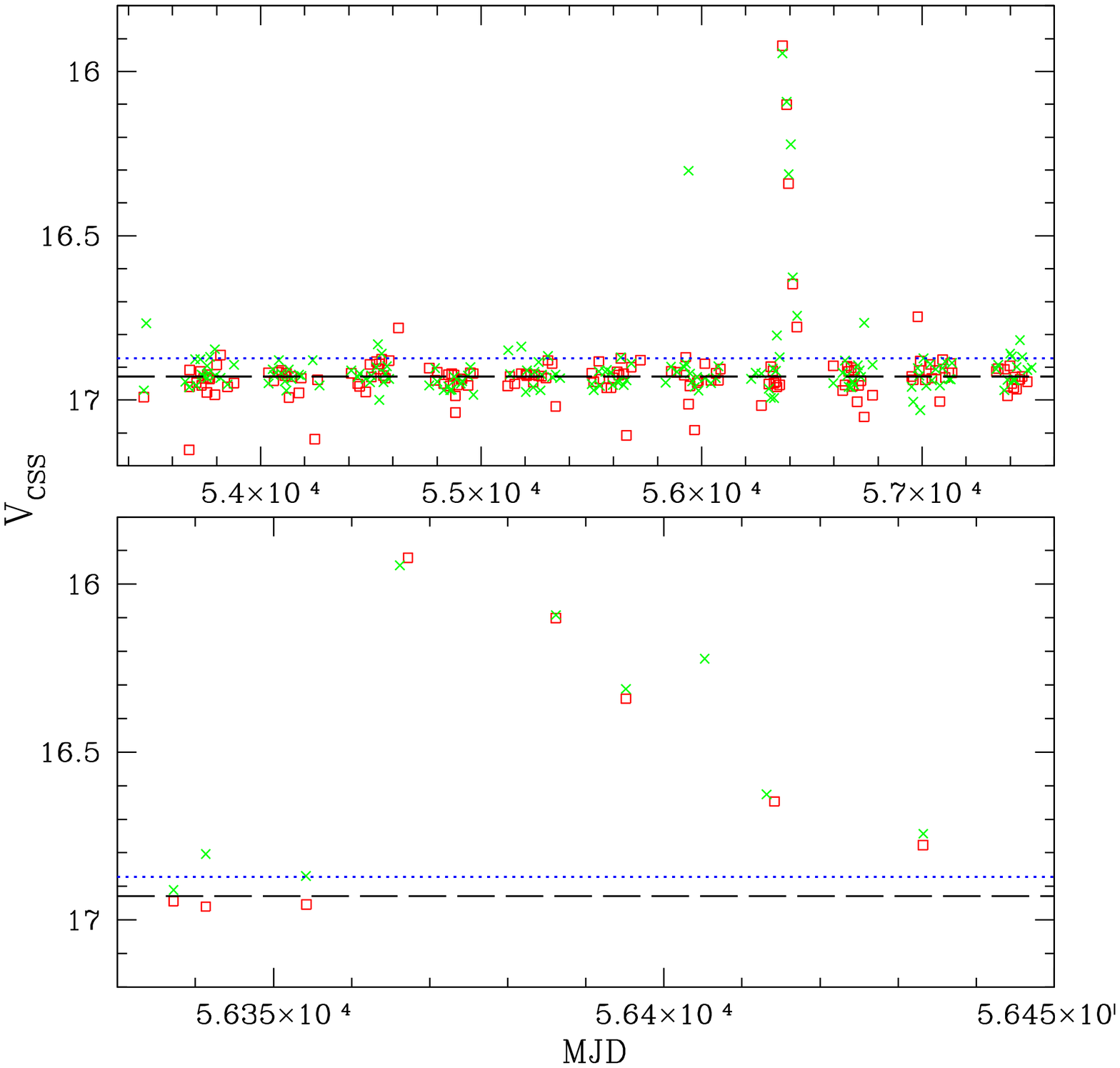}
\includegraphics[width=84mm]{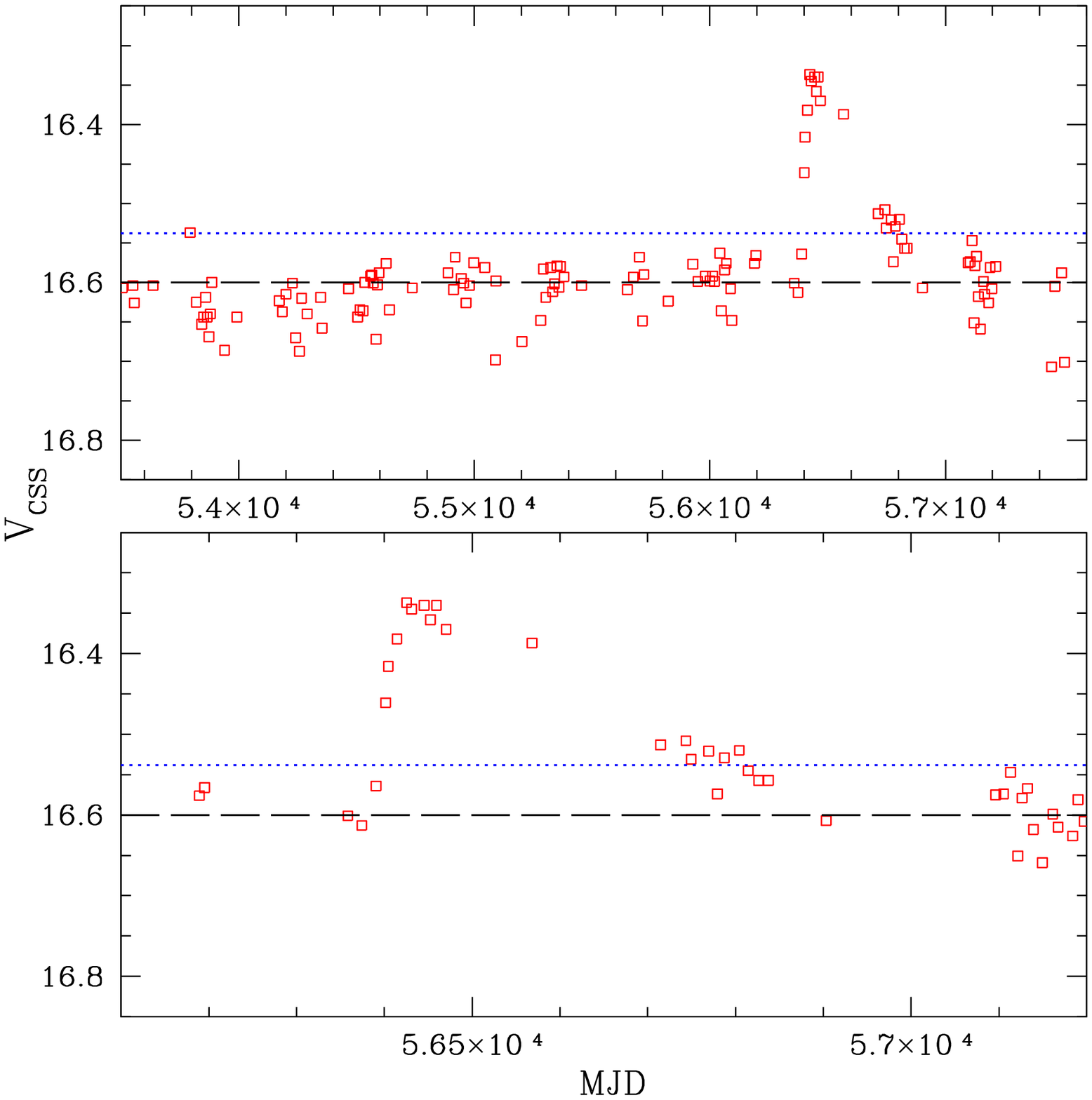}
\contcaption{
Nightly-averaged lightcurves for CRTS\_OBC\_454 (left) 
and CRTS\_OBC\_481 (right). CRTS\_OBC\_454 is in overlap
between two fields, providing an additional set of measurements 
shown as green crosses.
The host galaxies are KUG 0930+304 (aka SDSS J093332.30+301342.8, 
$z=0.027$) and SDSS J155045.39+201454.0 ($z=0.036$), respectively.
}
}
\end{figure*}

In Figure \ref{CRTS_OBC_284}, we present the full CSS lightcurves and a 
zoomed outburst region for a sample of the new outburst candidates that 
passed both lightcurve and image inspection. In each case we include the 
$1\sigma$ lower limit used to detect outbursts. We do not include
individual measurement uncertainties since our selection method is based on the distributions rather than individual uncertainties.

In most cases the events are seen to last around 100 days in the observed
frame. This timescale is completely consistent with expectations for
supernovae.  However, in a few cases the events are significantly
longer. For example, the outburst of CRTS\_OBC\_454 (Figure
\ref{CRTS_OBC_284}, set 4) appears to have lasted for at least 400 days. Such
events are not consistent with most types of supernovae. However, they
are consistent with some type-IIn supernovae that can have outbursts that
last for years (e.g. Smith et al.~2010, Rest et al.~2011).

\begin{figure}{
\includegraphics[width=84mm]{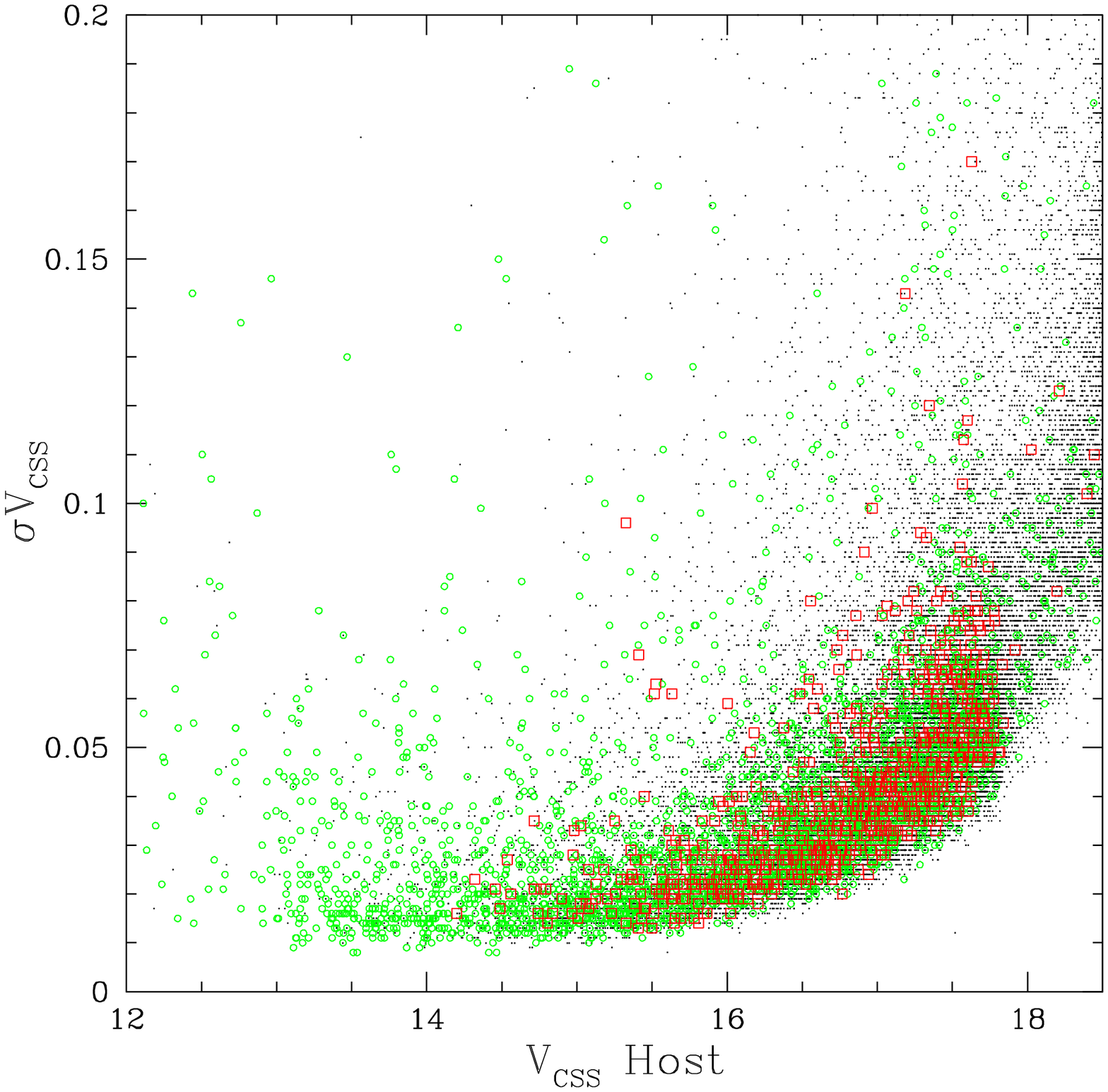}
\caption{\label{GalSSS}
The variability of new outburst candidates and previously known SN
candidates.  The large green dots show the scatter for galaxies with previously
known SN candidates. The red squares are the new outburst sources and the
black points are from a random sample of 50,000 SDSS galaxies with spectra.
}
}
\end{figure}

In Figure \ref{GalSSS}, we show the scatter of the newly selected outburst 
candidates compared to the previously known SN candidates. Interestingly, the 
new candidates are generally found to exhibit smaller photometric scatter than 
those galaxies with known SN candidates.
Also, as expected, the scatter among the new candidates is far smaller 
than the objects expected to be affected by nearby bright stars (as previously 
shown in Figure \ref{GalS}). That is to say, almost all of the new candidates 
have $\sigma < 0.2$ mag, while many of the galaxies affected by nearby saturated 
stars have $\sigma \sim0.8$ mag.

\begin{figure*}{
\includegraphics[width=84mm]{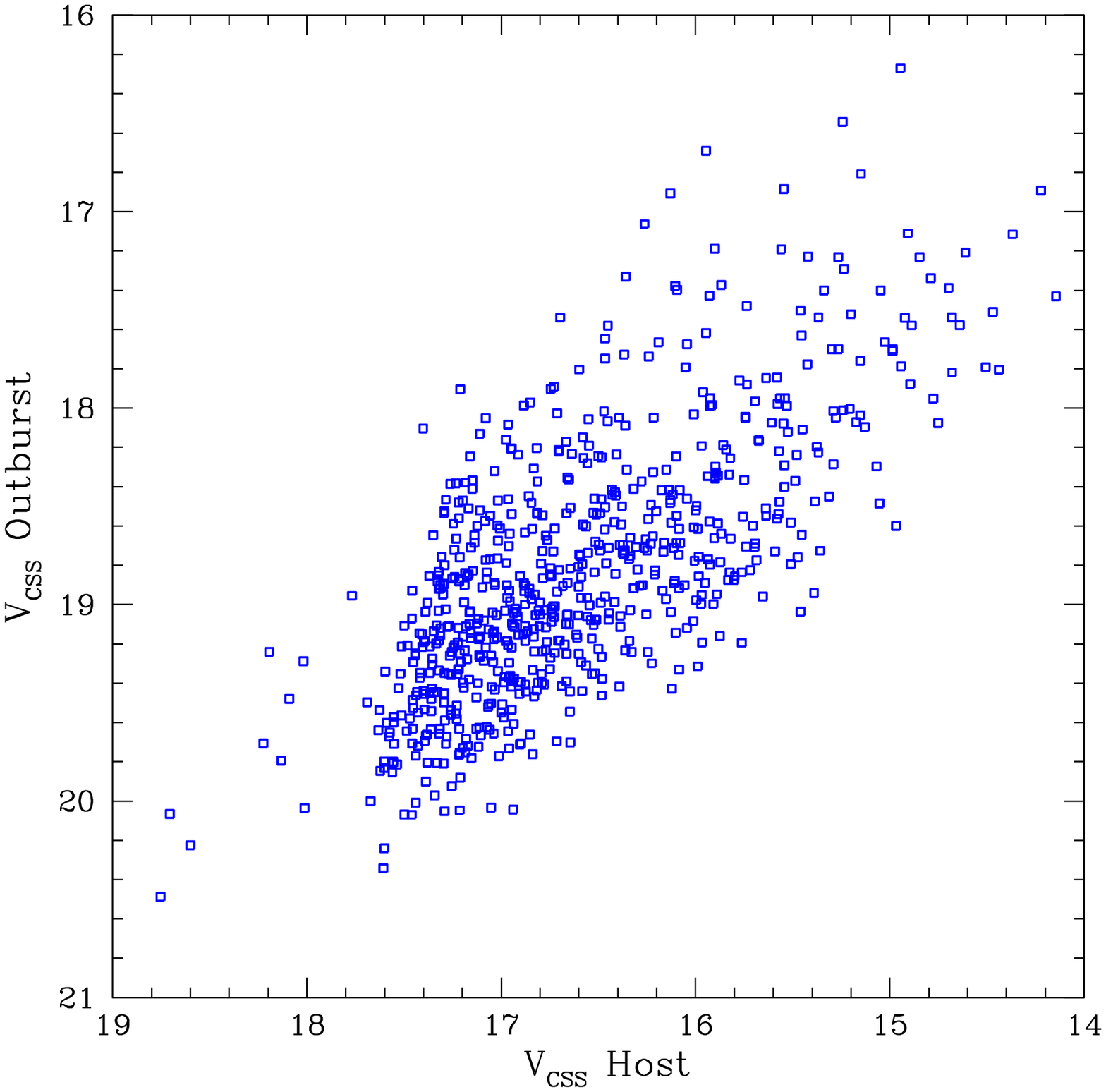}
\includegraphics[width=84mm]{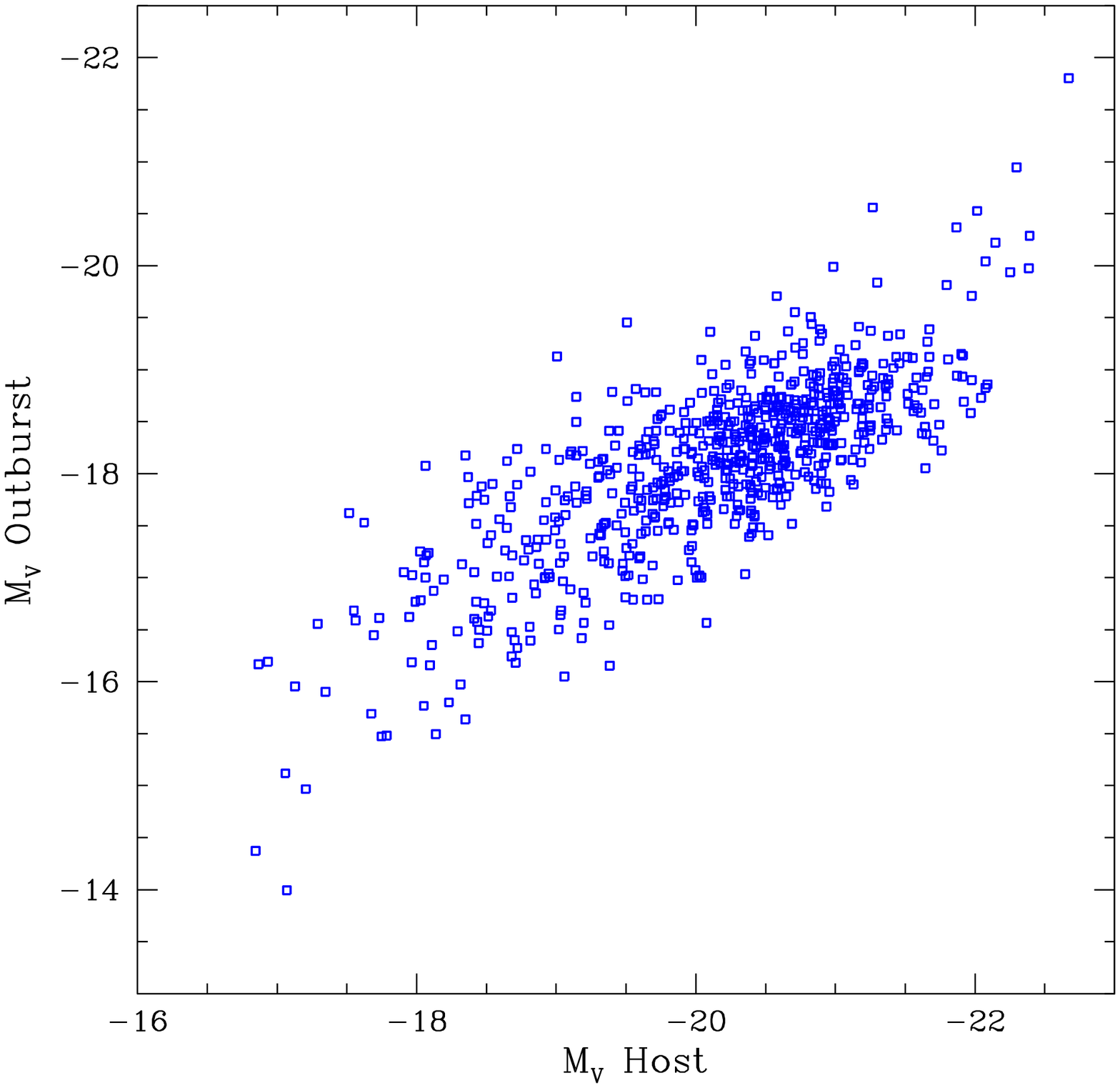}
\caption{\label{GalM}
Outburst and host magnitudes.  The left panel compares the apparent magnitudes
of the outburst sources with those of their hosts.  The right panel compares
the absolute magnitudes of the newly discovered outbursts with those of their
hosts based on their SDSS redshifts after correcting for foreground
extinction using the Schlegel et al.~(1998) reddening maps.
}
}
\end{figure*}

By subtracting the median host magnitudes from the peak observed magnitudes 
for each event, we derived estimates for the brightness of each outburst.
By using distances derived from the known host redshifts we also determined 
the peak absolute magnitude. However, these estimates are lower limits since, 
while we have corrected for foreground extinction (using Schlegel et al.~1998 
reddening maps), the amount of extinction the outbursts suffer within the hosts
is unknown. 

In Figure \ref{GalM}, we plot the magnitude distributions for the outbursts
and their hosts. As expected, the outburst absolute magnitudes generally
range from $M_V \sim -14$ to $-20$. This is completely consistent with the
brightness distribution for regular supernovae (Richardson et al.~2002).
However, there are a few brighter events that we will discuss later.
Overall, the outbursts are found to be asymmetrical. Those events where the
outburst shape is well defined all have much shorter rise times than
declines. Such features are a general feature of all types of SNe
lightcurves, even though there is significant diversity in the exact
shapes. In contrast, Graham et al.~(2018) found a several examples of QSO
flares where the rise and decline times were similar.

\section{Nature of the Host Galaxies}

\begin{table*}
\caption{Properties of outburst host galaxies\label{galtab}}
\begin{center}
\begin{tabular}{@{}crrcccrcrrc}
\hline
SDSS ID & RA (J2000) & Dec & $z$ & $V_{\rm CSS}$ & $M_V$ & $\rm MJD_{spec}$ & SFR  & Metallicity & Age & Mass\\
   & (deg) & (deg) &  & & & & ($log(\rm M_{\sun}.yr^{-1})$) & & (Gyr) & (log($\rm M_{\sun})$)\\
\hline
SDSS\_J002316.49-101850.2 &     5.8187 &   -10.3139 &  0.0282 & 16.10 & -19.31 & 52145 & 0 & 0.004 & 0.4042 & 8.64\\
SDSS\_J003042.78-085502.5 &     7.6782 &    -8.9174 &  0.1397 & 16.92 & -22.15 & 52146 & 0 & 0.01 & 1.1391 & 10.44\\
SDSS\_J003159.53-103646.0 &     7.9981 &   -10.6128 &  0.0545 & 16.04 & -20.85 & 52146 & 0.4 & 0.01 & 1.2781 & 9.79\\
SDSS\_J004310.40-090324.8 &    10.7933 &    -9.0569 &  0.0577 & 15.46 & -21.56 & 52162 & 0 & 0.02 & 4.75 & 10.75\\
SDSS\_J004701.14-085811.5 &    11.7547 &    -8.9698 &  0.1134 & 17.22 & -21.36 & 52148 & 0.9 & 0.02 & 1.2781 & 10.12\\
SDSS\_J004844.33-092643.3 &    12.1847 &    -9.4454 &  0.0958 & 16.95 & -21.23 & 52148 & 0 & 0.02 & 5.5 & 10.73\\
SDSS\_J005516.14+000542.0 &    13.8172 &     0.0950 &  0.0433 & 15.95 & -20.42 & 51783 & 0 & 0.02 & 4.25 & 10.22\\
SDSS\_J010921.02+154406.6 &    17.3376 &    15.7352 &  0.0603 & 17.15 & -19.97 & 51878 & 0.1 & 0.04 & 1.9 & 9.86\\
SDSS\_J011050.82+001153.3 &    17.7118 &     0.1981 &  0.0176 & 14.71 & -19.65 & 51816 & 0.2 & 0.01 & 1.2781 & 9.36\\
\hline
\end{tabular}
\end{center}
\begin{flushleft}
{\it The full table will be available online.}\\
Column 1: SDSS ID of the outburst host galaxies.\\
Columns 2 and 3: Source coordinates.\\
Column 4: Redshift of the galaxies from SDSS.\\
Columns 5 and 6: Apparent and absolute magnitudes
of the host based on CSS photometry after correcting for foreground 
extinction using the Schlegel et al.~(1998) reddening maps.\\
Column 7: Date that the SDSS spectrum was obtained.\\
Columns 8, 9, 10 and 11: Star formation rate, metallicity 
(model values 0.004, 0.01, 0.02, 0.04), age and host mass for the 
best fit model to the SDSS spectrum.\\
$^a$ denotes whether a galaxy is an AGN based on the spectrum and/or WISE.\\
\end{flushleft}
\end{table*}

SDSS DR13 provides details such as age, metallicity, starformation rate and 
mass for most of the galaxies with spectra. In Table \ref{galtab} we present 
the known properties for each of the galaxies containing an outburst candidate.

\begin{figure}{
\includegraphics[width=84mm]{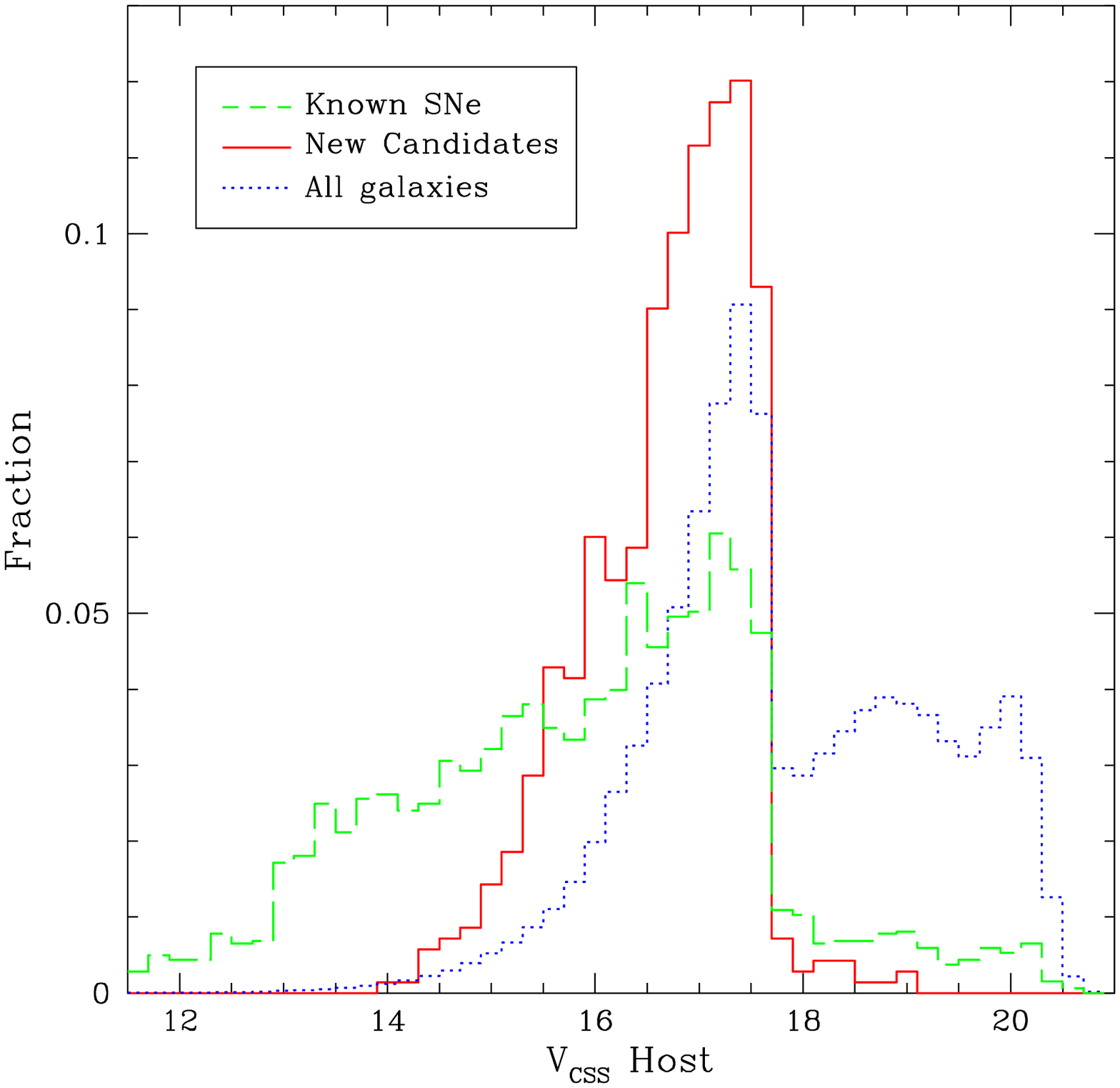}
\caption{\label{GalB}
The distribution of apparent $V_{\rm CSS}$ magnitude for SDSS galaxies.  The
blue-dotted line shows the distribution for the 1.4 million SDSS galaxies
with measured redshifts $z < 0.5$ and CSS photometry.  The green-dashed line
shows the distribution of redshifts for the 3,340 galaxies with known
SN candidates.  The red solid line shows the distribution for galaxies with newly
discovered outbursts.
}
}
\end{figure}

In Figure \ref{GalB}, we plot the distribution of the apparent magnitudes of
all SDSS galaxies with CSS matches along with those containing new CSS
outburst candidates.  The galaxies with new outburst candidates are clearly
peaked at bright magnitudes.  This peak coincides directly with that seen in
the distribution of SDSS galaxy magnitudes.  The strong peak in SDSS galaxy
magnitudes is due to the $r_P \le 17.7$ spectroscopic target selection.
Fainter spectroscopically confirmed SDSS galaxies come from the BOSS survey
as well as other later generation SDSS surveys. Our selection is clearly
not uncovering new outbursts in the fainter galaxies.
However, Figure \ref{GalB} does show that the hosts with new outburst 
candidates are generally fainter than those with previously known SN 
candidates. We believe this is predominately because the nearby 
bright galaxies are better monitored and the bright SNe that they 
contain have already been discovered. 

\begin{figure}{
\includegraphics[width=84mm]{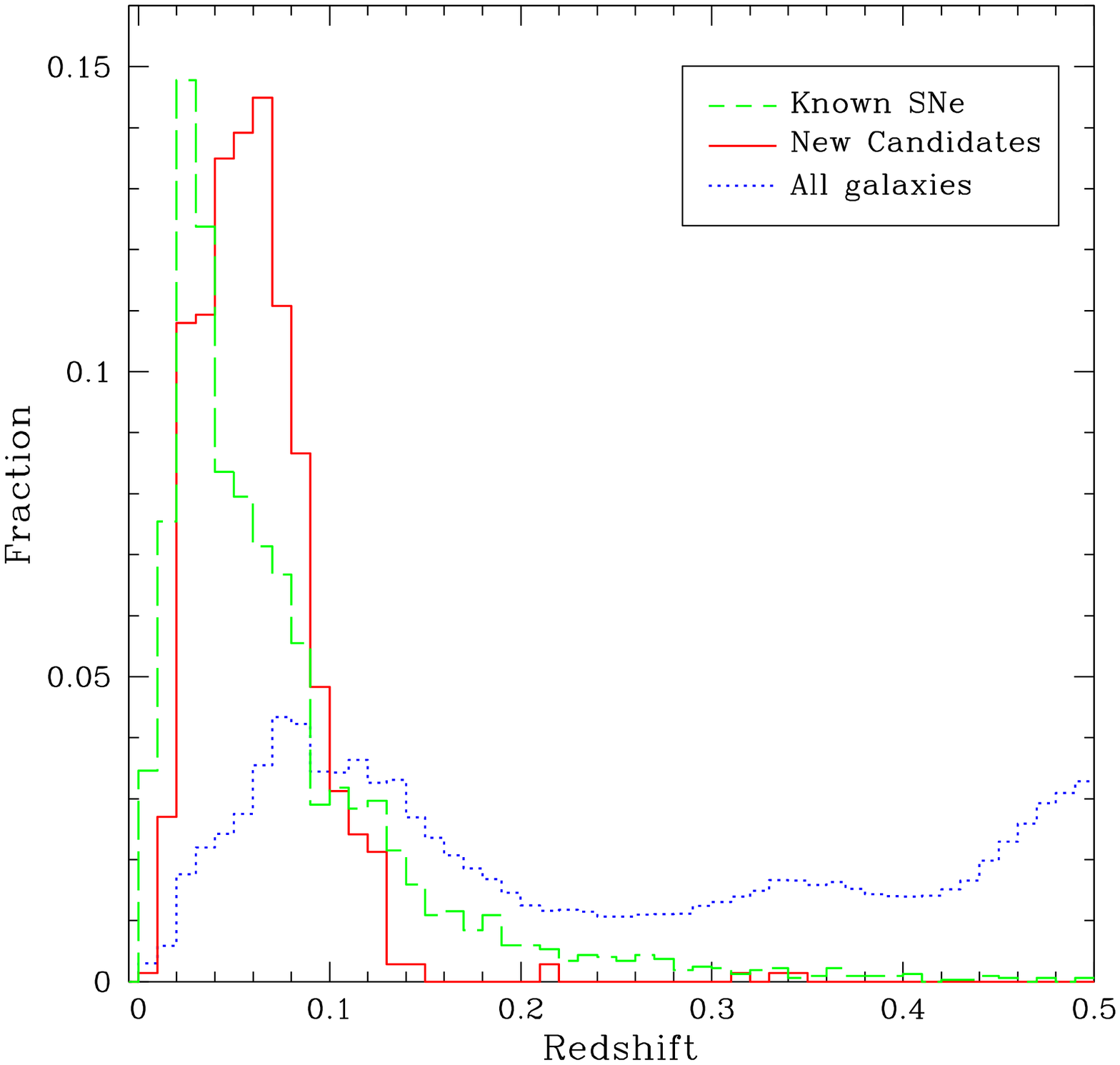}
\caption{\label{GalZ}
The distribution of SDSS galaxy redshifts. The blue-dotted line shows 
the distribution for the 1.4 million SDSS galaxies with measured redshifts 
$z < 0.5$ and CSS photometry.
The green-dashed line shows the distribution of redshifts 
for the 3,340 galaxies with known SN candidates.
The red solid line shows the distribution for galaxies with 
newly discovered outbursts.
}
}
\end{figure}

The redshift distributions of the 1.4 million matched SDSS galaxies is shown
in Figure \ref{GalZ}.  Here we note that the SDSS BOSS galaxies generally
have redshifts $z > 0.2$.  Thus, we find that the lack of new outburst
candidates in the faint SDSS BOSS galaxies relates to them being generally
more distant.  This agrees with the idea that most of the new outbursts are
due to supernovae. That is, SNe in general have peak magnitudes $M_V >
-19.3$ (the peak luminosity for type-Ia SNe). At $z=0.2$ regular SNe have
peaks with $V > 20.5$. Thus, such events are undetectable in the CSS data.
However, superluminous SNe can be much brighter and are detectable to
$z\sim0.3$.

\subsection{Emission line galaxies}

SDSS DR13 provides an assessment of the nature of
each spectrum by way of an assigned class. This parameter has values 
of ``STAR", ``QSO" and ``GALAXY", and is determined on the basis of 
the chi-squared value of the data with respect to a model (Bolton et al.~2012). 
Galaxies with narrow emission lines are sub-classified into AGN 
and star-forming types based upon measurements of $\rm [O{\scriptstyle\, III}]/H_\beta$ 
and $\rm [N{\scriptstyle\, II}]/H_\alpha$ line ratios. The resulting subclasses are noted 
as ``Starforming", ``Starburst" and ``AGN". An additional separate 
assessment of ``Broadline" is appended for spectra having emission 
lines with width $\sigma > 200\,{\rm km}\,{\rm sec}^{-1}$.

\begin{table}
\caption{Emission line galaxy subclasses\label{galprop}}
\begin{center}
\begin{tabular}{@{}lrrr}
\hline 
Subclass    & All \% (total)  & SN Hosts \% (total)  & Outbursters \% (total) \\
\hline
Starforming & 16.8 (238299)   &  47.0 (1474)  & 57.2 (396) \\
Starburst   & 4.4   (62829)   &  9.8  (308)   & 15.7 (109) \\
AGN         & 1.7   (24579)   &  2.7  (120)   & 2.3  (16) \\
Broadline   & 1.6   (22421)   &  7.0  (220)   & 1.0  (7) \\
\hline
\end{tabular}
\begin{flushleft}
Column 1: Subclasses defined by SDSS.\\
Column 2: Number of emission line galaxies in
each subclass as fraction of the full 1.4 million sample.\\
Column 3: Fraction of all SN hosts in the 
respective SDSS subclass.\\
Column 4: Number of new outbursters in each
respect SDSS subclass.
\end{flushleft}
\end{center}
\end{table}

In Table \ref{galprop}, we present the percentage of the total sample for each 
of the galaxy subclasses having emission lines.  Interestingly, while only $\sim16.7\%$
of the 1.4 million SDSS are classified as starforming, 47\% of the known SNe
hosts and 57\% of the hosts with newly outbursts are starforming. 
Higher numbers of CCSNe are expected in starforming galaxies than 
passive ones since they are due to massive stars with short lives. 
In contrast, type-Ia supernovae are known to occur in galaxies 
regardless of recent star formation. 
Nevertheless, the fact that we observed $3.4$ times as many new outburst
candidates in starforming galaxies is larger than expected since our sample
is mainly magnitude-limited by the SDSS legacy survey $r_P \le 17.7$
follow-up limit.  For that limit Li et al.~(2011a) show that the number of
type-Ia's far exceeds CCSNe, since type-Ia's are significantly brighter on average.
In contrast, Li et al.~(2011a) found that for their volume-limited sample,
$\sim 76\%$ of their SNe were core-collapse.

An important difference between our galaxy sample and that of Li et
al.~(2011a,2011b), is that the latter may have missed a larger fraction 
of nearby faint dwarf galaxies since most such galaxies were uncatalogued 
at the time their survey began.  Thus, it is very unlikely that their 
volume-limited sample was as complete as ours at the low-mass end. Dwarf galaxies 
often have very high specific star formation rates (SSFR, Li et al.~2011b, 
Huang et al.~2012). 
Furthermore, our selection of candidates from lightcurves naturally introduces 
a bias toward detecting outbursts in intrinsically faint galaxies (since the 
variation in such a galaxy's lightcurve will be larger). This effect is balanced 
slightly by the fact that fainter local galaxies are missing from the SDSS 
spectroscopic sample because of the magnitude limit.
Indeed, dwarf galaxies do not seem to play a significant role since the new 
outburst hosts generally have $-18 < M_V < -22$, and are thus as much as $\sim 5$ magnitudes brighter 
than the faintest dwarf galaxies where SNe have been discovered (Williams et al.~2008, 
Drake et al.~2011b).

\begin{figure}{
\includegraphics[width=84mm]{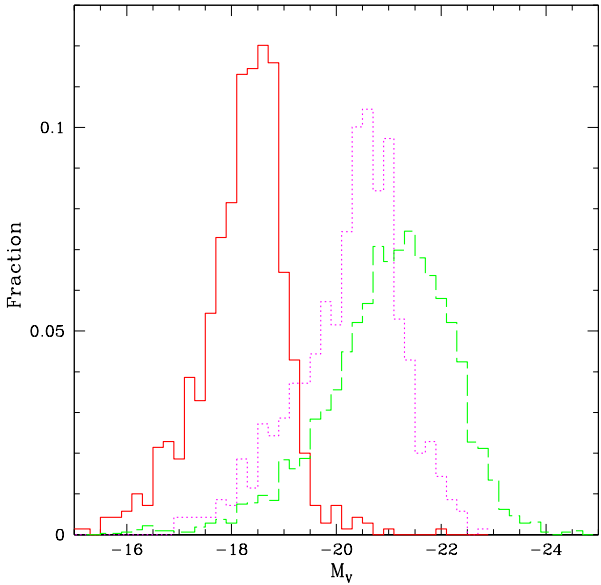}
\caption{\label{Absmag}
The relative distributions of absolute brightness of host galaxies.
The dashed-green histogram gives the previously discovered SN candidate hosts. 
The dotted-magenta histogram gives the magnitude distribution of the 
host galaxies with newly discovered outbursts. The solid-red histogram gives 
the distribution of the outburst absolute magnitudes after subtracting 
the host galaxy fluxes.
}
}
\end{figure}

As a more complete test, we consider the combination of the known and new outburst 
hosts since many of the SNe in intrinsically bright galaxies were already known. 
Figure \ref{Absmag}, shows that the host galaxies of the new outbursts 
are $\sim 1$ mag less luminous (on average) than those with previously known SN 
candidates. In the combined known plus new outburst case we see that 
49\% of the galaxies are starforming.
When we add the starburst host galaxies to the starforming ones we see 
that $\sim 60\%$ of the new outburst plus known SN candidates are in active 
starforming galaxies. Thus, overall most of host galaxies with outbursts are 
undergoing significant star formation.

We also note that the number of new outbursts associated with
broad-lined hosts is much lower than those of previous SN candidates.  This
may suggest that some of the previously discovered and unconfirmed SN
candidates were due to AGN activity.  It is also possible that some of the
galaxies contain broad lines due to supernovae rather than a super-massive
black hole (SMBH). Indeed, a number of SNe have already been found within
the SDSS spectroscopic galaxy sample (Graur et al.~2015). However, significant
contamination seems unlikely since spectra with SN represent
$\sim 0.01\%$ of the SDSS galaxies analysed by Graur et al.~(2015).  

Although only a small fraction of the new outbursts are associated
with hosts that are noted by SDSS as either AGN or having broad lines, 
it is worth considering AGN as a possible source for the new events. 
In particular, it is possible that a number of AGN have not been 
identified due to the weakness of visible narrow lines (Koss et al.~2017).

\subsubsection{Active Galactic Nuclei}

AGN are responsible for the bulk of the optical variability observed 
in galaxies. Their presence is almost guaranteed when searching
for variability within galaxies. However, the nature of AGN variability 
itself is known to be quite well modeled as a damped random walk process
(Kelly, Bechtold, \& Siemiginowska 2009, MacLeod et al. 2012).
Flaring events that appear to be associated with AGN have only been found 
in the last decade as samples of hundreds of thousands of AGN are 
regularly monitored by transient surveys (MacLeod et al.~2016, Graham et al.~2017). 
In some cases these events are believed to be due to phenomena
such as microlensing, superluminous SNe, tidal disruption 
events (e.g., Lawrence et al.~2016), or changes in the accretion or 
dust environment that might cause an AGN to appear to turn 
on (e.g., LaMassa et al.~2015, Gezari et al.~2017, Stern et al.~2017).

\begin{figure*}{
\includegraphics[width=84mm]{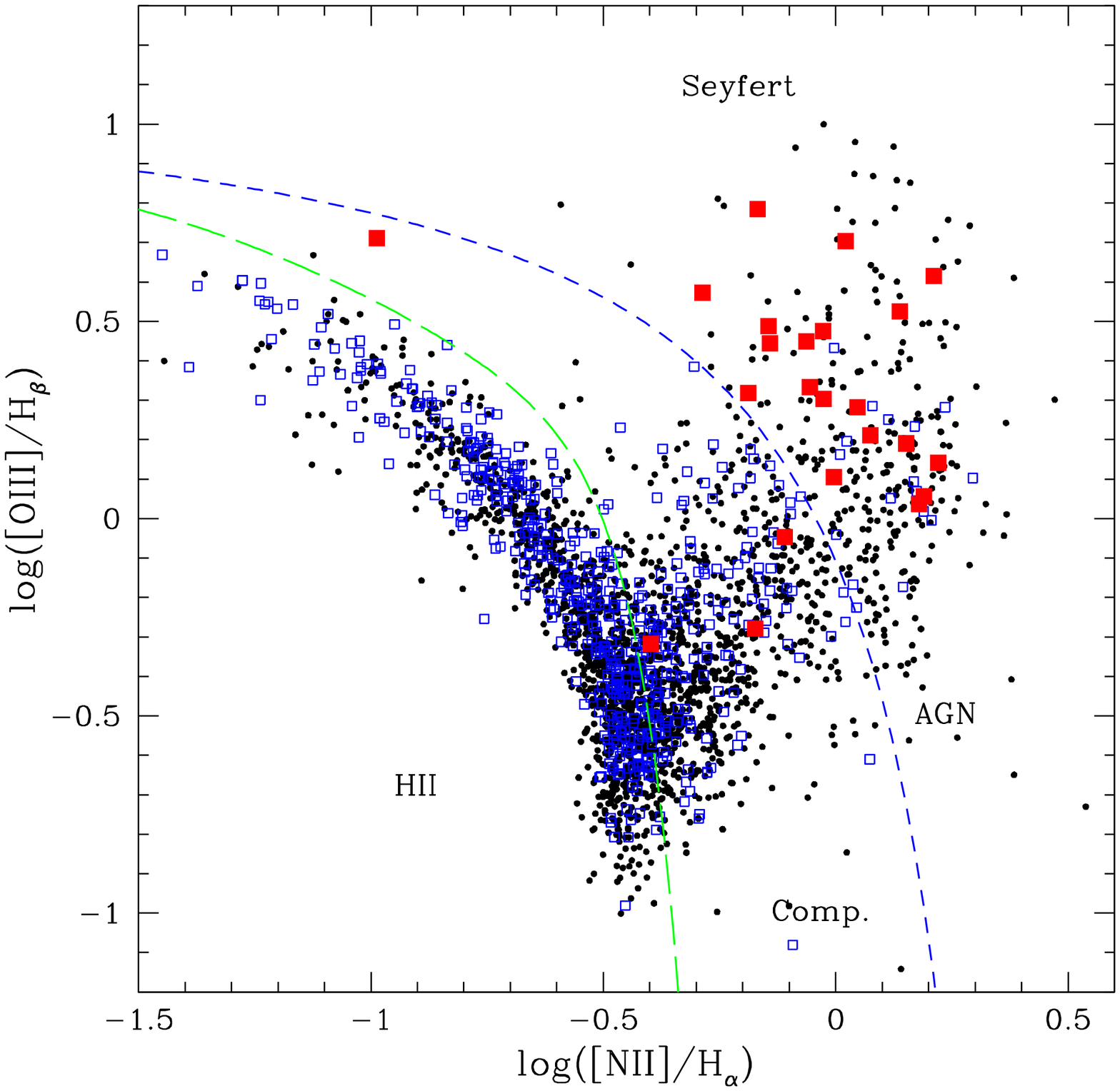}
\includegraphics[width=84mm]{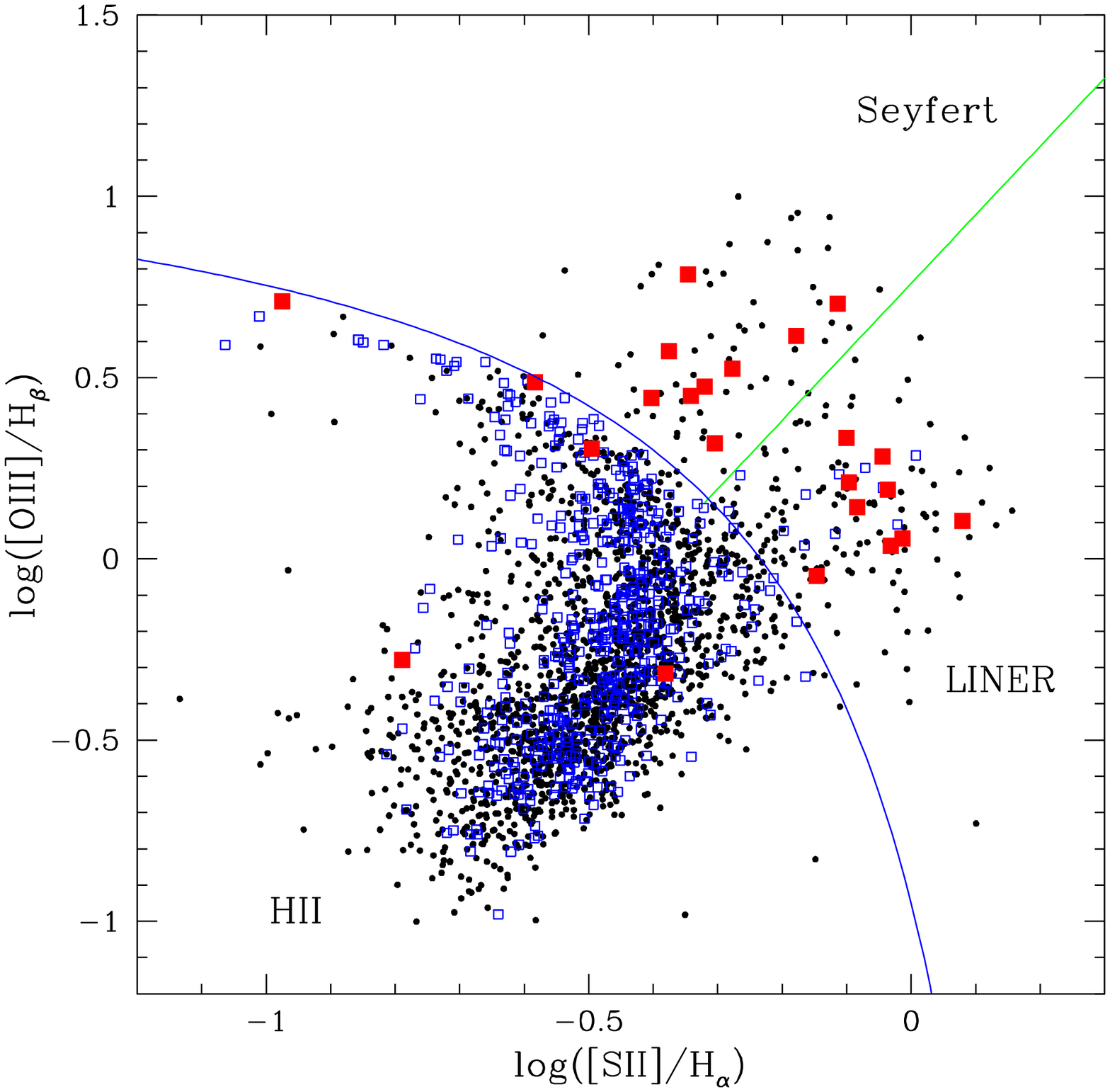}
\caption{\label{Emiss1}
Emission line diagnostics for previously known 
SN candidate host galaxies and new outburst hosts.
The small dots show the known SN hosts
while the open blue boxes show the new outburst 
hosts and filled red squares are known AGN.
In the left panel we divide the galaxies 
into AGN and those that are dominated by star formation 
(i.e. {H}{2} galaxies). Between these types of objects there 
are composites of the two. In the right panel we separate
the starforming galaxies from Seyfert and LINER type AGN.
}
}
\end{figure*}

\begin{figure*}{
\includegraphics[width=84mm]{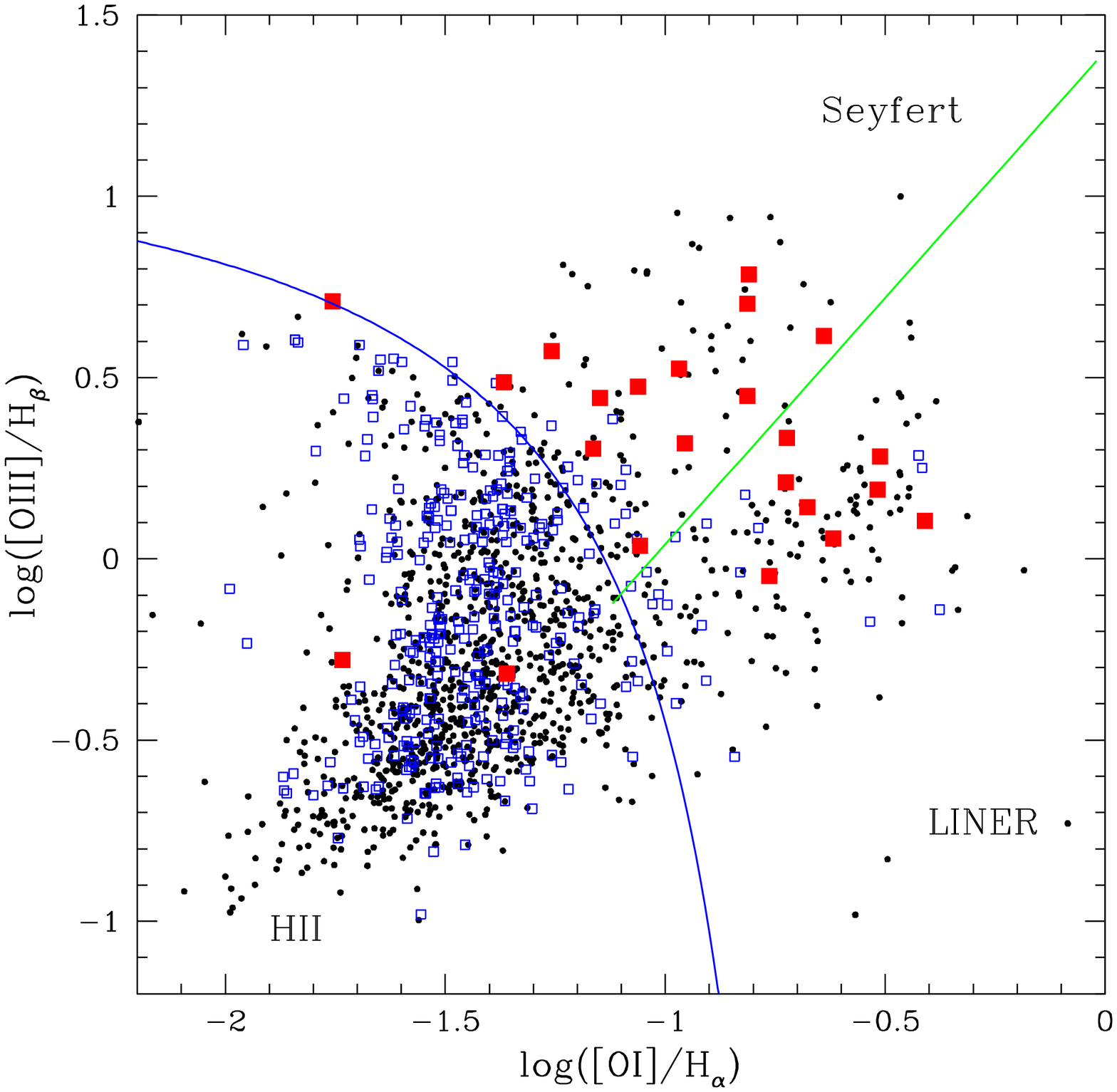}
\includegraphics[width=84mm]{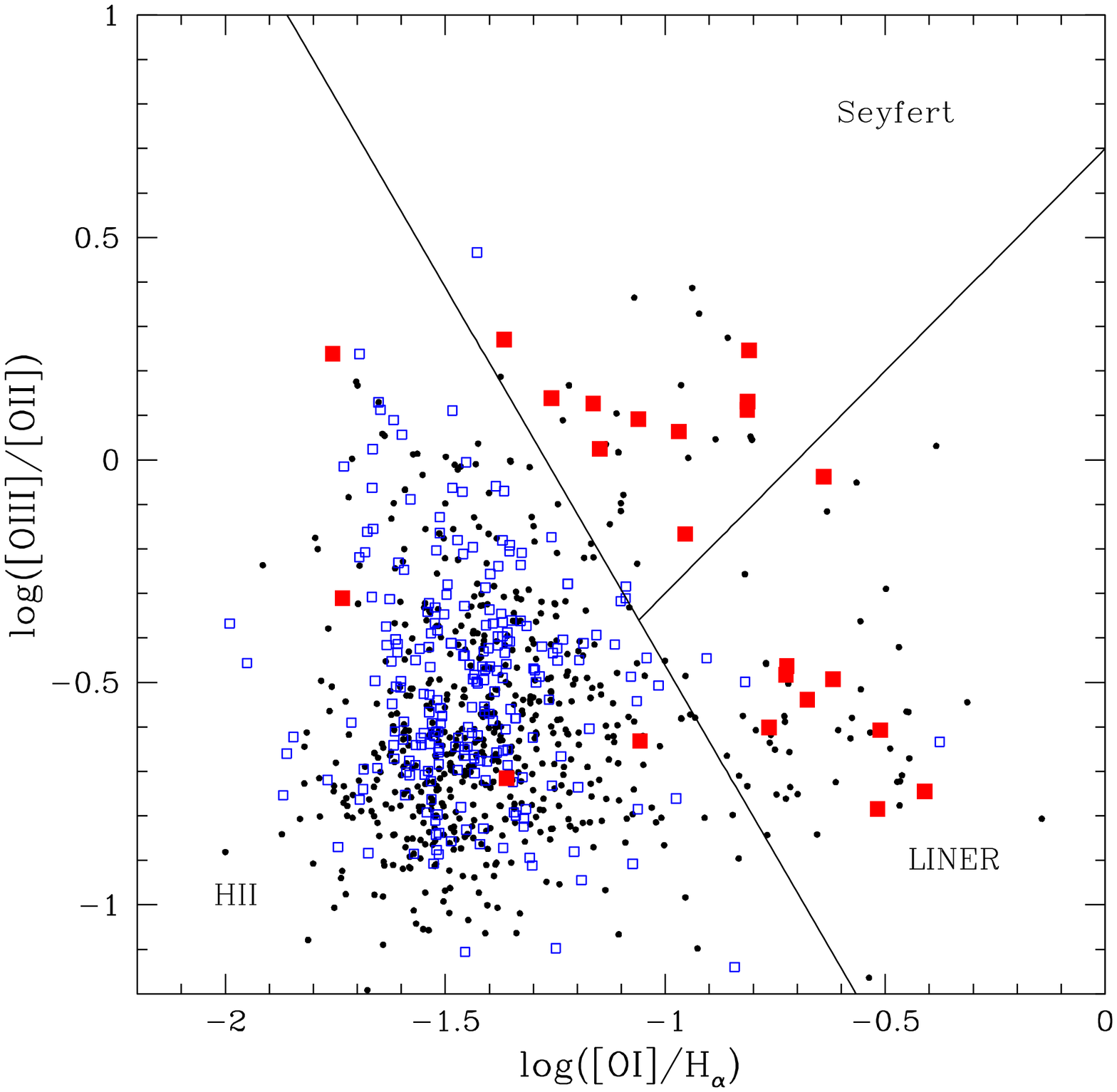}
\caption{\label{Emiss2}
Emission line diagnostics as in Figure \ref{Emiss1},
but using $\rm [O{\scriptstyle\, I}]$ lines 
rather than $\rm [S{\scriptstyle\, II}]$ and
$\rm [N{\scriptstyle\, II}]$.
In both panels we divide the starforming galaxies 
from Seyfert and LINER AGN. Fewer galaxies are 
present than in Figure \ref{Emiss1} due to the 
generally lower strengths of $\rm [O{\scriptstyle\, I}]$.
}
}
\end{figure*}

To understand how many of the newly detected outbursts might be 
due to AGN variability, we decided to investigate the nature of 
the host galaxies. The SDSS AGN classifications are based solely 
on the ratios of $\rm [O{\scriptstyle\, III}]$, $H_\alpha$ and
$\rm [N{\scriptstyle\, II}]$ lines.
Since the selection of AGN via a single set of criteria is subject 
to uncertainties and the presence of individual lines and accuracy of 
their measurements, we decided to calculate the four sets of line 
ratio diagnostics given by Kewley et al.~(2006) for each galaxy with 
narrow emission lines. 
The four diagnostic diagrams are presented in Figures \ref{Emiss1} and \ref{Emiss2}
for the hosts of both previously known SN candidates and new outburst candidates.
In addition to the lines used by SDSS, these include measurements 
for $\rm [S{\scriptstyle\, II}]$, $\rm [O{\scriptstyle\, III}]$ 
and $\rm [O{\scriptstyle\, II}]$. Such line diagnostics have been 
widely adopted for the purpose of AGN classification.

The measurements we use are based on Thomas et al.~(2013) and come 
from SDSS DR13, where they are known as the {\it Portsmouth} values. 
SDSS DR13 provides line measurements for each galaxy with a spectrum. 
However, while inspecting the fits to the SDSS spectra we found that some 
of the emission line measurements provided by SDSS DR13 were spurious. In 
particular, we found that galaxies lacking emission lines still included
as positive values with very large relative errors. Thus, rather than 
simply adopting the SDSS fit values, we limited our selections to galaxies 
where the measured fluxes were at least twice their assigned uncertainties. 
This reduces the number of galaxies with new outburst candidates with
line measurements by $\sim50\%$ in the case of the generally weak 
$\rm [O{\scriptstyle\, I}]$ line as seen in Figure \ref{Emiss2}.

Since many of the galaxies do not have measurements passing our criteria, we
decide to only require that each AGN candidate met three of the four AGN
diagnostics. Furthermore, since some AGN with noisy spectra may not meet even
this selection, we also inspected each spectrum for the presence of broad
lines that would also indicate the presence of an AGN (rather than pure star
formation).  As expected, we found very good agreement between the objects
we selected as AGN and those noted as AGN in SDSS DR13.

Clearly a small number of the galaxies with outbursts are indeed classified 
as AGN based on emission line diagnostics. Yet overall the bulk of the 
galaxies are not AGN, but are instead star-forming. Nevertheless, a number of the hosts also 
fall in the composite AGN-starforming region shown in Figure \ref{Emiss1}.
This indicates that the number of AGN identified in this way is only a lower limit. 
Many additional AGN could be highly obscured and thus not exhibit strong emission 
lines. We do not observe any significant difference in the distribution of the hosts 
of our new outburst candidates and those of previously known SN candidates.

In addition to investigating line diagnostics, we also matched our 
candidates with AGN candidates selected based on AllWISE IR data 
by Secrest et al.~(2015). This catalog consists of $\sim 1.4$ million 
AGN candidates. Only four of the host galaxies with new outbursts 
had matches with this catalog.

In total, 23 of the galaxies with outbursts were found to be in hosts that
were likely AGNs. These galaxies are marked in Table \ref{galtab}.  It is
quite clear that only a small number of the outburst candidates can be
attributed to either regular stochastic AGN variability or rare AGN
flares.  Of course, SNe also occur in galaxies with AGN. So the fact that a
candidate occurs in a host with an AGN does not mean that an outburst is due
to an AGN. Indeed, the rate of SNe in galaxies with AGN is expected to be
similar to that in non-AGNs (Cappellaro, Evans, \& Turatto~1999).  Since our
sample of outbursting events in AGN is so small. It is not possible to judge
whether the AGN with broad lines (type-1 AGNs) is different than the rate
in type-2's where the broad lines are obscured. Nevertheless, the presence 
of AGN variability would certainly increase the threshold required to detect 
outbursts in AGN. Thus SNe are expected to be under-represented in AGN

\subsubsection{Starforming Galaxies}

The rate of CCSNe is expected to be highest in galaxies with active star
formation (Petrosian et al.~2005). This clearly is borne out by the very
high fraction of new outbursts that we found to be in starforming
galaxies. To better understand this link, we extracted masses and star
formation rates for each galaxy with values given by DR13, based on Maraston
et al.~(2006).  This set consists of values for $\sim 1$ million of the
galaxies in the full sample of SDSS galaxies. Thus, only $\sim 30\%$ of the
galaxies in our CSS-SDSS sample do not have star formation measurements.

In addition to star formation rates, the SDSS galaxies also have mass
determinations.  In order to have reasonable confidence in the mass
estimates, which are based on fits to the SDSS spectra, we decided to 
only include those hosts where the goodness-of-fit value was 
$\chi_r^2 < 1.5$.

\begin{figure*}{
\includegraphics[width=84mm]{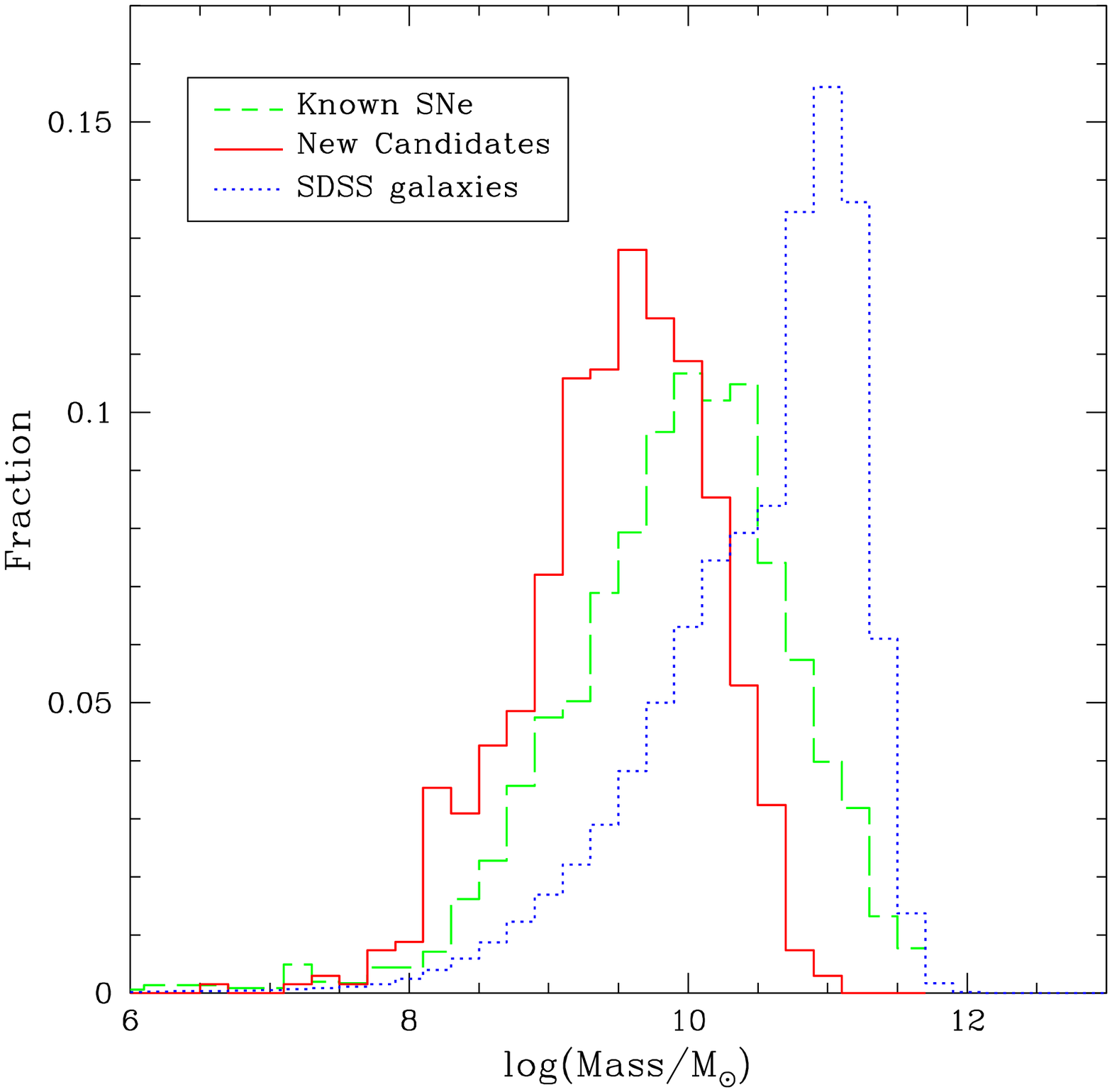}
\includegraphics[width=84mm]{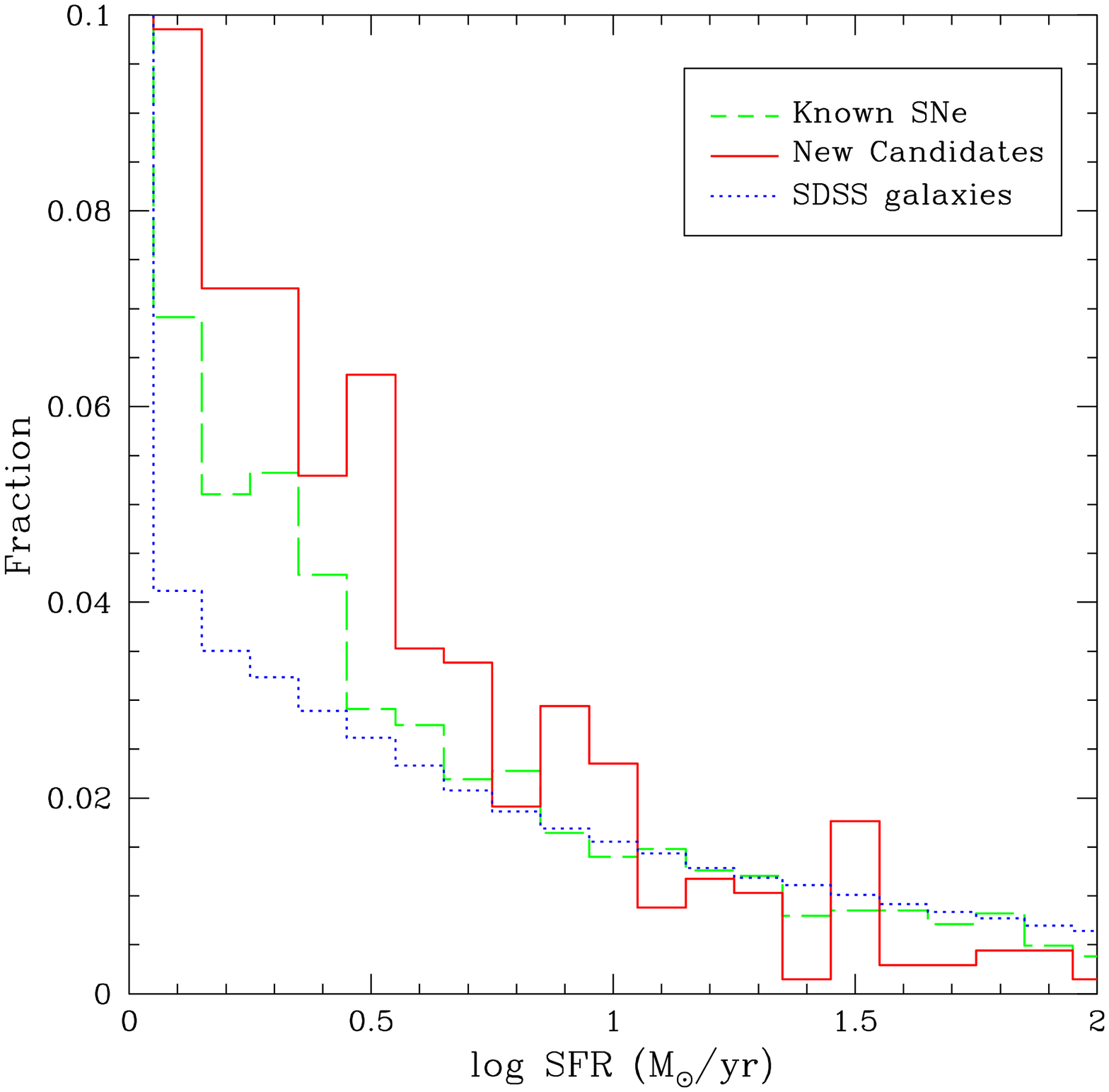}
\caption{\label{Mass}
SDSS galaxy properties.  The left panel shows the distribution of
galaxy masses. The blue-dotted histogram gives the distribution for 1.0
million SDSS galaxies with mass estimates and goodness-of-fit $\chi^2_r < 1.5$
from SDSS DR13. 
The dashed-green line shows the distribution for SDSS hosts with previously 
known SN candidates. The solid-red line shows the distribution for  
new outburst candidates. The right panel shows the star-formation
rates for the same groups of galaxies (using the same colour coding).
}
}
\end{figure*}

In Figure \ref{Mass}, we plot the host masses and star formation rates for 
the full sample of SDSS galaxies as well as those with previously known SN
candidates and the new outbursts. 
Overall, we see that the new outbursts have lower mass hosts than the
previously known candidates and the overall galaxy population.  This effect 
is most likely due to the fact that the low-mass hosts are less luminous. 
Thus, the outbursts can be more easily detected. Additionally, as noted above, 
many low-mass hosts have high star formation rates.

As expected, most of the galaxies in the full SDSS sample have low star
formation rates. We see that the hosts of new outbursts and previously known
SN candidates generally have higher rates than the full sample. This is
consistent with the high fraction of events in starforming galaxies noted
above. These starforming galaxies tend to have rates of 0.1 to 0.6 $\rm
M_{\sun} yr^{-1}$.  Intriguingly, the fraction of galaxies with $\rm SFR >
10 M_{\sun} yr^{-1}$ is the same for the full SDSS galaxy sample as for the
hosts of known SN candidates ($\sim15\%$).  It is not clear is this a real
effect, or inaccurate star formation rates due to the presence of flux from
contaminants (such as AGN).

\section{Event Timescales}

As noted earlier, the bulk of the new outbursts are almost certainly due 
to SNe. In cases where a new source is seen offset from the core of a
galaxy, this is almost a certainty. However, strong evidence for the 
SN origin of the events also comes from their timescales.

For each outburst candidate we measured the timespan over which the 
event was continuously $>1\sigma$ brighter than the median. Since 
the outbursts are not very well sampled in general, this provides 
only a rough estimate for the timescale for each event. 

\begin{figure}{
\includegraphics[width=84mm]{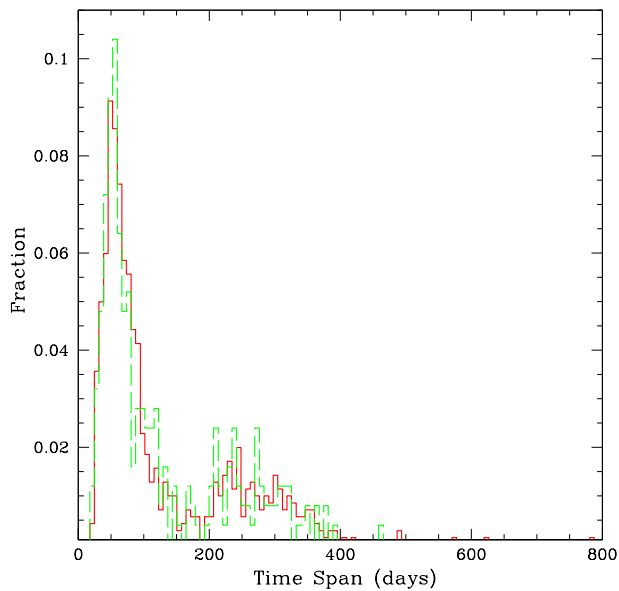}
\caption{\label{Timescale}
Event timescales.
The timespan distribution for known SN candidates 
is plotted with the dashed-green histogram. New outburst
candidates are plotted as the solid-red histogram.
}
}
\end{figure}

Figure \ref{Timescale} shows the distribution of outburst 
timescales along with the timescales for the previously known 
SN candidates that are present in the CSS photometry.  Clearly, 
the two timescale distributions are an excellent match, and 
both are strongly peaked near a time span of 50 days.

\begin{figure*}{
\includegraphics[width=84mm]{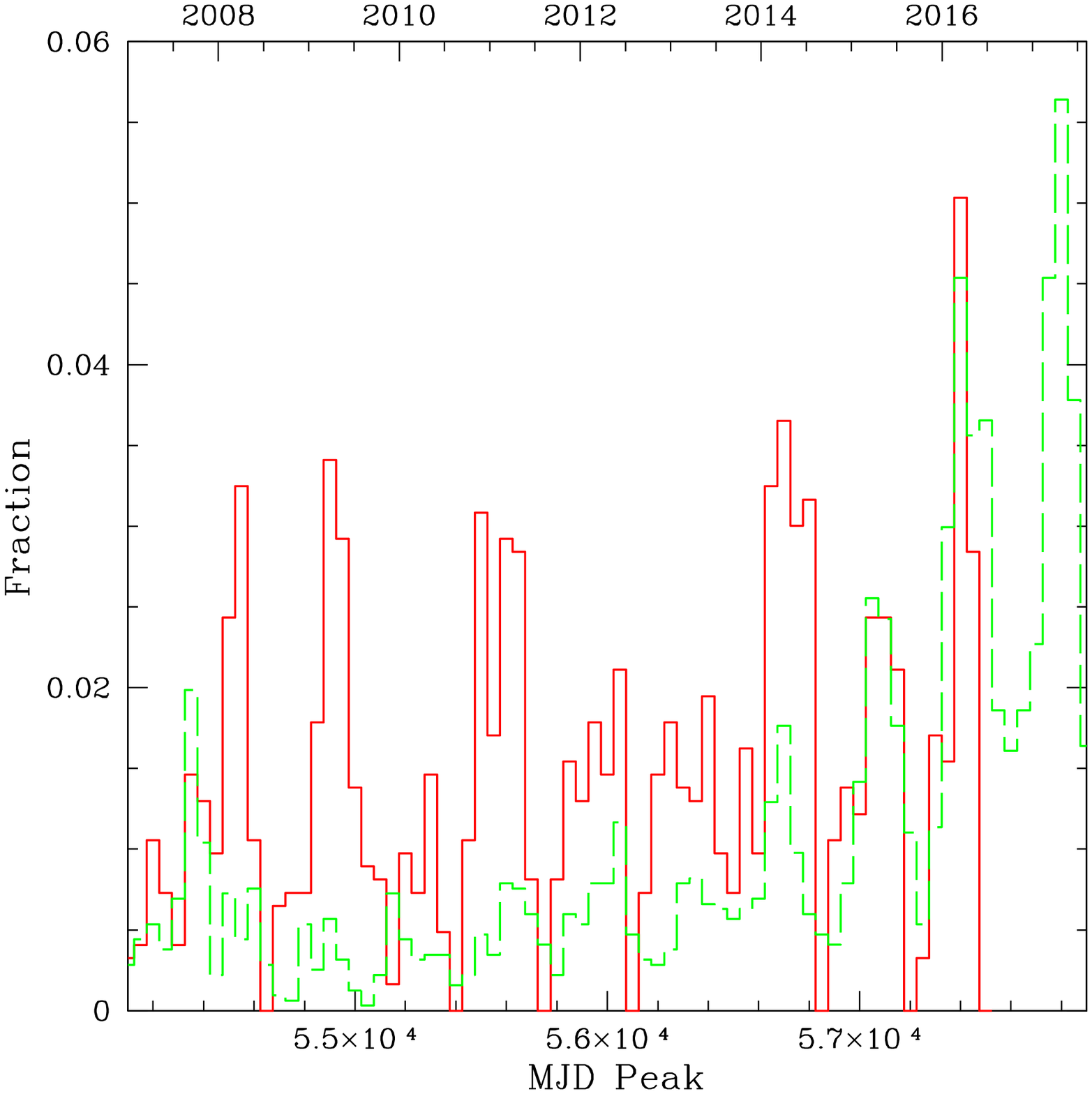}
\includegraphics[width=84mm]{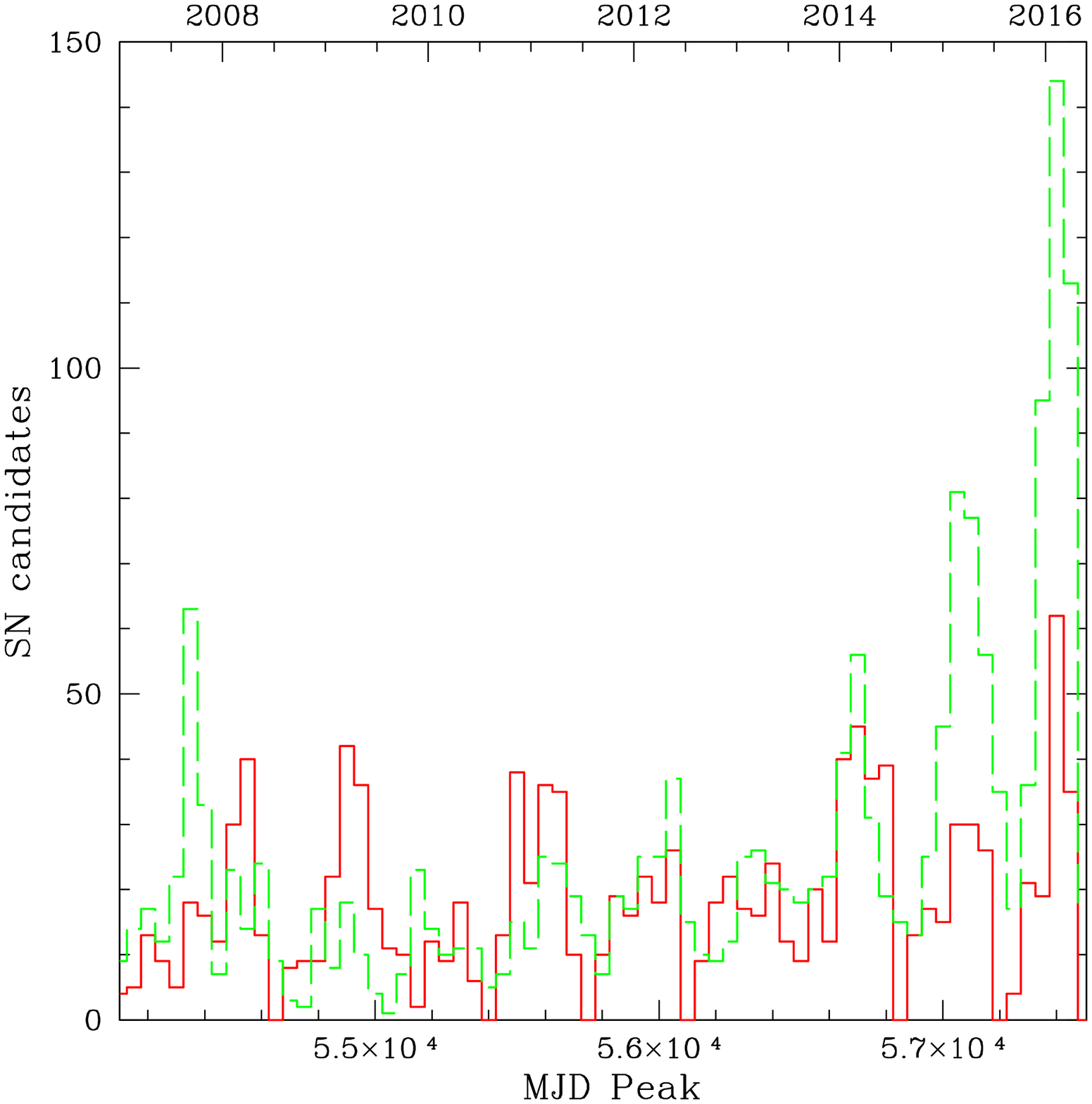}
\caption{\label{Dates}
The peak times of outbursts.
In the left panel we show the temporal distribution 
of outburst detections. The red solid-line shows the 
fraction of newly detected sources, with the green 
dashed-line showing the distribution for previously 
known SN candidates.
The right panel shows the actual number of detections in 
bins spanning 50 days (using same colour coding as the left panel).
}
}
\end{figure*}

In Figure \ref{Dates}, we show the temporal distribution of event peaks for
both previously known SN candidates and new outbursts. This figure shows that 
the new detections are generally clustered in time (due to the fact that the 
SDSS fields are only visible for part of the year). It also shows that the 
rate of SN discovery, in spectroscopically confirmed SDSS galaxies, has 
increased dramatically in over the past few years.

In general, we expect the true timespan of the events to be longer
since sparse sampling and observing gaps evident in Figure \ref{Dates}
often cause the beginning or end of an event to be missed. In addition 
to the main peak, the timescale distributions exhibit a second broad 
feature.

The longest events in the timescale distributions are inconsistent with 
those of most common SNe. That is to say, regular SNe typically last 
for around one to two hundred days and have rise times of
less than $\sim 40$ days.  Longer timescale types of SNe consist of 
type II-Ps and type IIns.  Type II-Ps are often detectable for a couple
of hundred days (Li et al.~2011a), while in exceptional cases type-IIns 
can last for many years (Rest et al.~2011, Fox et al.~2015).  However, both
type-IIPs and long-timescale IIns have been found to comprise $< 10\%$ of
SNe in the Li et al.~(2011a) magnitude-limited SN sample, although 
II-Ps comprise 40\% of SNe in a volume-limited sample (Li et al.~2011a).  
Note however that some of the outburst timescales are likely errant due 
to uncertain start and end times.

In contrast to our outburst candidates, Graham et al.~(2017) found 51 large
amplitude ($\delta V_{\rm CSS} > 0.5$ mag) outburst events with typical timescales
of 900 days from a sample of 900,000 QSOs.  In our analysis we only find a
few long timescale events in our sample of 1.4 million galaxies. This
strongly suggests a real difference in the number of long timescale
events between normal galaxies and AGN, since
only $\sim 3\%$ of our input galaxy sample are known AGN. Nevertheless, 
it is worth considering how other selection effects
might caused a difference in the observed timescales.

For the bulk of events that occur in galaxies that lacking evidence for an 
AGN, it is unlikely that our selection process misses events lasting hundreds
of days with amplitudes $>0.5$ mags. This is because the scatter in 
the lightcurves is generally $< 0.5$ mags and most lightcurves 
span more than 3000 days. However, events lasting a few
thousand days could be missed since the event itself would 
increase the photometric scatter and thus in increase the detection 
threshold. In the case of AGNs, events could also be missed due to the 
increased scatter from intrinsic variability. Searches based on lightcurve 
fits, such as carried out in Graham et al.~(2017), are preferable for 
such sources.

\section{MLS Sources}

Of the $\sim$1.7 million SDSS DR13 galaxies with spectra and $z < 0.5$,
$\sim$380,000 overlap with the Catalina MLS survey. Since MLS has a much
smaller field-of-view than CSS, the coverage is sparse away from the
ecliptic. Only about half (152,000) of the matches between MLS and SDSS have
20 or more nights of observations.  Thus, the full sample of MLS-SDSS
matches is only $\sim10\%$ that of the CSS set. 

Applying the same selection criteria as with the CSS data, we found 561
initial candidates. After inspecting the lightcurves, and removing the
previously known SNe candidates, this number is significantly reduced to just 
70 candidates.  We then inspected 400 image cutouts associated with the peaks 
detected for these candidates.  This reduced the number of MLS outburst 
candidates to just 38.

\begin{table*}
\caption{Properties of MLS outbursts\label{mlsout}}
\begin{minipage}{186mm}
\begin{tabular}{@{}crrccrrrrrc}
\hline
ID & RA (J2000, deg) & Dec (deg) & $V$ & $M_V$ & $\rm MJD_{peak}$ & length & Signif & log(P) & Nights & Quality\\
\hline                                                     
CRTS\_OBM\_1 &     0.8805 &     1.2243 & 19.39 & -17.90 & 55130.2 &  399 & 36.00 & -15.39 & 4 & I$^b$\\
CRTS\_OBM\_2 &     1.8685 &     0.7770 & 19.44 & -17.69 & 55090.4 &   68 & 17.20 & -12.66 & 5 & I \\
CRTS\_OBM\_3 &    19.3826 &     8.7039 & 20.87 & -17.07 & 54760.2 &  371 & 12.38 & -12.36 & 4 & I \\
CRTS\_OBM\_4 &    26.7793 &    13.9415 & 19.31 & -17.67 & 55486.2 &  367 & 33.27 & -9.75 & 3 & I$^a$ \\          
CRTS\_OBM\_5 &   116.1534 &    22.3044 & 19.93 & -17.98 & 55601.2 &   82 & 11.11 & -9.09 & 3 & I$^a$$^b$\\
CRTS\_OBM\_6 &   117.7806 &    16.6848 & 18.91 & -17.94 & 54833.5 &   89 & 18.66 & -16.28 & 6 & I$^a$ \\
CRTS\_OBM\_7 &   118.4133 &    20.9226 & 19.97 & -18.73 & 56188.5 &  361 & 36.29 & -33.04 & 14 & I$^b$\\
CRTS\_OBM\_8 &   119.2773 &    21.2713 & 19.76 & -18.86 & 55578.3 &   38 & 18.23 & -16.68 & 5 & I $^a$ \\
CRTS\_OBM\_9 &   120.2549 &    21.1292 & 20.47 & -18.03 & 55602.2 &   37 & 12.00 & -9.31 & 3 & I  $^a$$^b$\\
CRTS\_OBM\_10 &   122.5167 &    18.9427 & 19.68 & -16.78 & 55505.5 &   96 & 10.61 & -9.83 & 4 & I  $^a$ \\
CRTS\_OBM\_11 &   128.4806 &    17.4051 & 20.12 & -18.18 & 56213.5 &   84 & 16.31 & -11.58 & 4 & I  $^a$$^b$\\
CRTS\_OBM\_12 &   131.2395 &    19.1262 & 19.71 & -18.15 & 54138.3 &  245 & 24.55 & -12.04 & 4 & I  $^a$ \\
CRTS\_OBM\_13 &   138.2021 &    15.4437 & 20.05 & -18.79 & 55988.2 &   78 & 13.32 & -10.97 & 4 & II \\
CRTS\_OBM\_14 &   141.8882 &    13.8358 & 20.07 & -17.44 & 55981.2 &  232 & 19.17 & -13.05 & 4 & I$^a$$^b$\\
CRTS\_OBM\_15 &   150.3220 &    15.1010 & 19.02 & -18.15 & 54530.3 &  300 & 16.77 & -12.00 & 3 & I$^a$ \\
CRTS\_OBM\_16 &   151.7186 &    10.9784 & 19.72 & -19.09 & 56371.2 &   72 & 12.90 & -11.63 & 3 & I$^a$ \\
CRTS\_OBM\_17 &   155.2287 &     9.8550 & 19.90 & -18.37 & 56297.5 &   42 &  8.72 & -8.80 & 3 & III  \\
CRTS\_OBM\_18 &   159.5800 &     6.7474 & 18.71 & -18.67 & 54554.3 &  218 & 27.87 & -18.69 & 5 & I$^a$ \\
CRTS\_OBM\_19 &   161.7155 &     6.6385 & 19.48 & -17.92 & 55893.5 &  272 & 17.92 & -10.21 & 4 & I$^b$\\
CRTS\_OBM\_20 &   167.8902 &     7.3996 & 19.74 & -17.37 & 55981.4 &  320 & 21.51 & -16.95 & 5 & I$^a$  \\
CRTS\_OBM\_21 &   168.9025 &     5.7472 & 19.59 & -18.45 & 57162.2 &  440 & 25.57 & -21.65 & 10 & I$^a$  \\
CRTS\_OBM\_22 &   170.8991 &     3.7309 & 21.08 & -20.62 & 55649.2 &  242 & 13.54 & -10.89 & 3 & II  \\
CRTS\_OBM\_23 &   175.9182 &     1.5518 & 19.51 & -18.10 & 55648.3 &   61 & 13.04 & -11.72 & 3 & I$^a$$^b$\\
CRTS\_OBM\_24 &   177.7692 &     0.2482 & 19.03 & -17.52 & 54197.3 &  227 & 11.03 & -10.37 & 4 & I$^a$ \\
CRTS\_OBM\_25 &   181.8560 &     0.8423 & 20.05 & -16.61 & 55327.2 &  301 & 41.73 & -12.00 & 3 & II \\
\hline
\end{tabular}
\end{minipage}\\
\begin{flushleft}
Column 1: CRTS outburst candidate IDs.\\
Columns 2 and 3: Source coordinates.\\
Columns 4 and 5: Apparent and absolute magnitudes
for the peak of the outburst, respectively.\\
Columns 6: MJD of the outburst peak.\\
Column 7: Timespan over which the candidate outburst 
was detected about $1\sigma$.\\
Column 8: Total significance in sigma of the detections
during the outburst timespan.\\
Column 9: Probability of false detection assuming normally
distributed data.\\
Column 10: Number of nights when the outburst was detected
above $1\sigma$.\\
Column 11: An assessment of quality of the outburst candidate 
based the inspection of lightcurves and images as well as the 
presence of detections in CSS data. The values are as follows:
$\rm{I}$ --- high confidence events, $\rm{II}$ --- moderate confidence 
events, $\rm{III}$ --- low confidence events.\\
$^a$ outburst is also detected in CSS observations.\\ 
$^b$ source is resolved from host galaxy.\\ 
$^c$ long timescale event.\\ 
\end{flushleft}
\end{table*}

The MLS data has much better resolution than CSS data since the original 
camera had $1\arcsec$ pixels (compared to $2.5\arcsec$).  For 13 of the 
38 outburst candidates we noted the presence of new source offset from the core 
of the host galaxy in the MLS images. These events are all likely to 
be SNe. Of the 38 candidates, 12 had already been found in our analysis of the 
CSS data. The remaining 25 new MLS candidates are presented in Table \ref{mlsout}.

Since the CSS survey covers all of the area of MLS, we extracted and
inspected the CSS lightcurves for each of the MLS outburst candidates. Among
the 25 new MLS outbursts, 16 showed some brightening in CSS data taken
during the event. Thus, these 16 events are all considered to be very strong
outburst candidates. The other MLS candidates generally lacked good CSS
coverage during the events.

As with CSS candidates, we extracted the lightcurves of all of the
previously known SNe candidates in the MLS-SDSS sample. Inspecting the MLS
lightcurves we found five events that were visible in the MLS data, yet
missed initial selection due to low significance. Five more were missed
since their lightcurves actually exhibited a decrease in flux during the
outburst. This decrease is due to the aperture of the SN sequestering 
some of the host galaxy's flux when it is detected as a separate source 
offset from the host.

Comparison of the rate of new outbursts per galaxy for the MLS data and
CSS data shows that the rate for MLS is only half that of CSS. 
This difference is due to two factors. Firstly, the MLS data has far fewer
observations per object on average than CSS. Specifically, during our 
analysis period the MLS dataset has an average of 43 nights of observations 
per galaxy, while CSS has 99.
Secondly, the MLS has higher resolution and better average observed seeing 
than in CSS. Thus any SNe that are offset from their hosts are more
likely to be detected as a new source in MLS data than in CSS data. 
This means that the MLS galaxy lightcurves are less likely to include
the flux of an offset SNe.

\section{Events of Interest}

\subsection{CRTS\_OBC\_204}

Although almost all of the new outbursts detected in CSS and MLS
are likely to be regular SNe, there are a few that clearly are not.
One of the unusual events we discovered in our analysis is CRTS\_OBC\_204,
occurring in the confirmed AGN containing galaxy SDSS J233454.07+145712.8.

\begin{figure}{
\includegraphics[width=84mm]{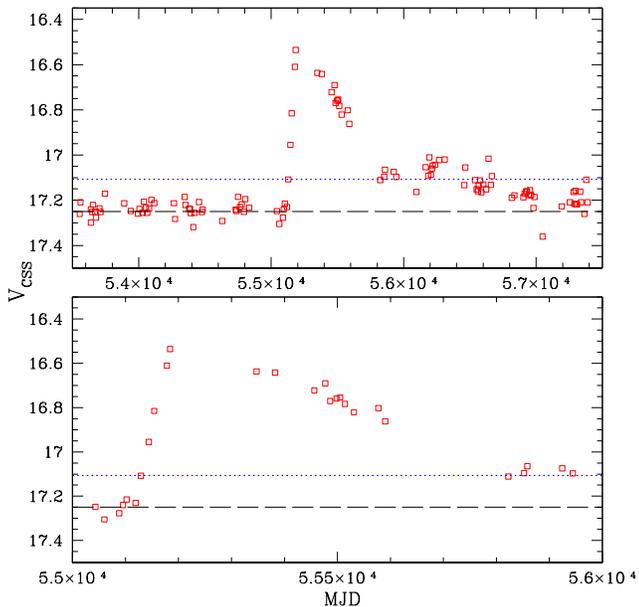}
\caption{\label{CRTS_OBC_204}
The lightcurve of outburst CRTS\_OBC\_204, 
occurring within the spectroscopically confirmed 
AGN SDSS J233454.07+145712.8.
}
}
\end{figure}

The lightcurve for this galaxy is presented in Figure \ref{CRTS_OBC_204}.
The source exhibits a very strong outburst that appears to 
have taken $\sim60$ days to reach a peak of $M_V = -21.1$ 
(after subtracting flux from the galaxy) in mid-December 2009.
Overall, this event appears to have lasted at least 1,700 rest 
frame days. The shape of the lightcurve is inconsistent with
that of any known type of SN. However, the peak brightness
is consistent with superluminous SNe. The host for this
event is clearly extended in SDSS images.

A low signal-to-noise ratio ($\sim3.5\sigma$) radio source is seen in the
NVSS survey (Condon et al.~1998) at the location of the host. However, the
flux appears extended and the source does not appear in the NVSS catalog.
This suggests that the flux may simply be due to noise. Furthermore, no radio
source is seen in VLSS (Perley et al.~2007) or TGSS survey data (Intema et
al.~2017).

The lightcurve of CRTS\_OBC\_204 is similar to that of CRTS transient
CSS120103:002748-055559, which was noted by Breedt, Gaensicke, \&
Parsons~(2012) as being due to a blazar at $z=0.428$. However, 
Breedt et al.~(2012) based their blazar classification on 
the detection of broad lines and a radio detection in NVSS data.
Our own analysis of NVSS data suggests that the significance of 
the radio detection is $< 2\sigma$.
Furthermore, CSS120103:002748-055559 does not appear in deeper 
FIRST survey radio data (Becker, White \& Helfand~1995), nor 
in TGSS or VLSS radio data. We suggests that neither CRTS\_OBC\_204,
nor CSS120103:002748-055559, are blazar outbursts. Rather they 
are flares from radio-quiet QSOs. The lightcurve of CRTS\_OBC\_204 
also resembles the QSO flares found by Graham et al.~(2017).

We obtained follow-up spectra of CRTS\_OBC\_204 with the Palomar 5m and
Double Beam Spectrograph (DBSP) on UT 2017 July 27 and 2017 August 29.
The new spectra were not significantly different from the original
SDSS spectrum. However, since our spectra were both taken many years after 
the peak of the event, when the source appears to have returned to quiescence,
this is not unexpected.

While this paper was in preparation CRTS\_OBC\_204 was identified as 
a potential AGN flare by Kankare et al.~(2017) based on the shorter 
lightcurves available in our public CSDR2 release.

\subsection{CRTS\_OBC\_62}

\begin{figure}{
\includegraphics[width=84mm]{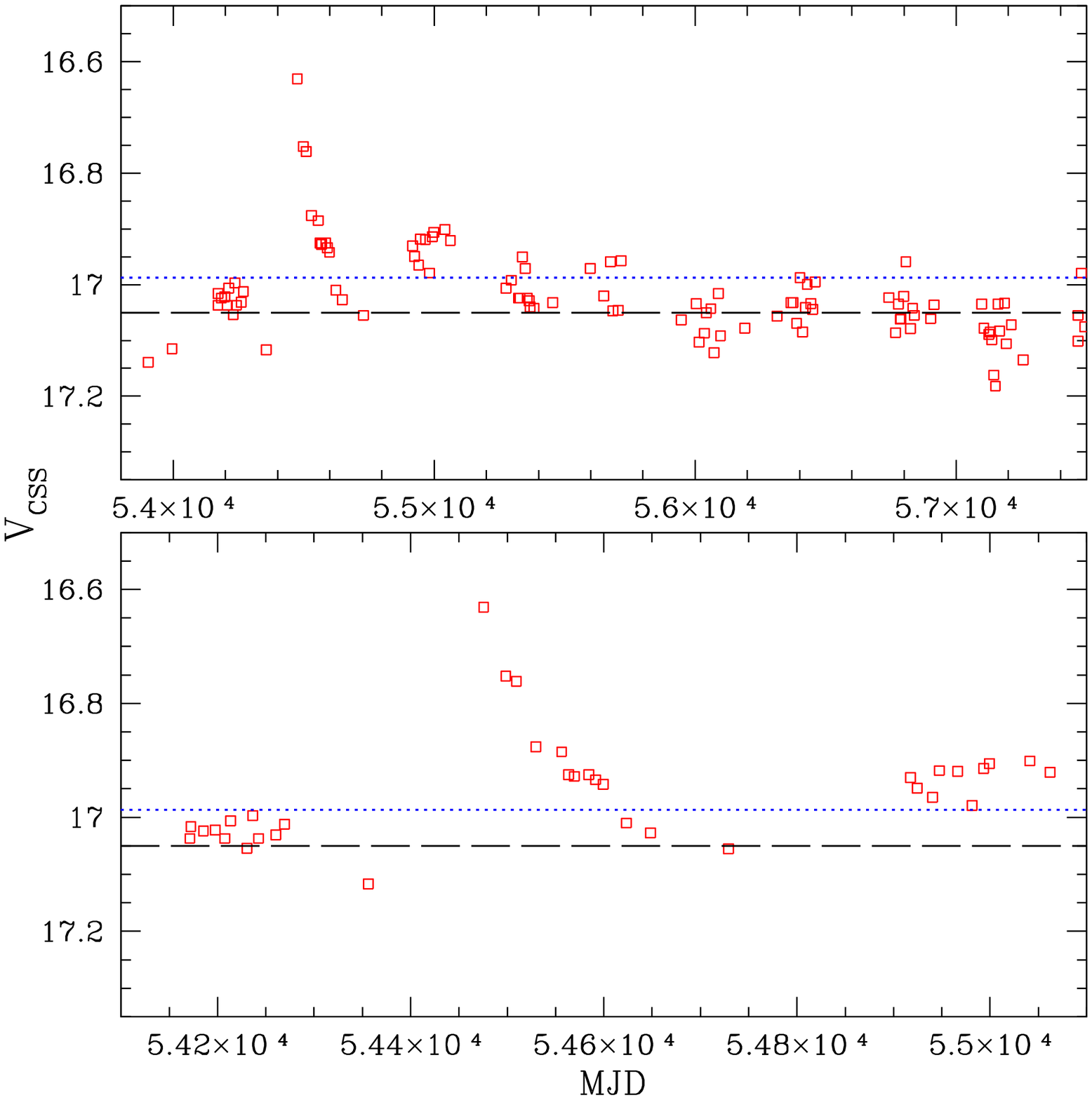}
\caption{\label{CRTS62}
The lightcurve of long timescale outburst CRTS\_OBC\_62,
occurring in blue dwarf galaxy C1543+091 (aka SDSSJ154543.55+085801.3).
}
}
\end{figure}

In Figure \ref{CRTS62}, we present the lightcurve of the long-timescale
outburst candidate CRTS\_OBC\_62 ($\alpha=\,$15:45:43.55, $\delta=\,$08:58:01.3 
J2000).  This event was found in known Lyman alpha
emitting blue dwarf galaxy 1543+091 (Meier \& Terlevich 1981).  As with
CRTS\_OBC\_204, after peak this event initially fades and then rises again
slightly to form a plateau that lasts for hundreds of days.

Interestingly, Izotov, Thuan \& Privon~(2012) noted that the host galaxy
displays the presence of $\rm [Fe{\scriptstyle\, V}]_{4227}$ and $\rm
[Ne{\scriptstyle\, V}]_{3426}$ which they attribute to hard ionizing
radiation. They also note that, although the spectrum shows the presence of
$\rm [He{\scriptstyle\, II}]_{ 4686}$, it lacks the bump signature associated
with Wolf-Reyet (WR) stars. They suggest that the source of the hard
radiation is either an AGN or fast radiative shocks, such as caused
by SNe.

Sartori et al.~(2015) selected this galaxy as a very rare ($< 1\%$ of 
their 50,000 galaxy sample) example of an AGN in a low mass galaxy 
($\rm log(M_{\sun})=8.2$) based on mid-IR data. However, Senchya et al.~(2017) 
suggest that such hard-ionizing spectral features with the lack of a 
WR bump are consistent with massive O-stars in a low metallicity 
environment (as is observed in 1543+091). 
Given the unclear nature of both the event and the galaxy, 
this object is worthy of further study.

\subsection{Other interesting events}

A number of the new outburst candidates in this analysis were found in
the same hosts as previously known SN candidates. For example, CRTS\_OBC\_9 
is a long event that occurred in 2007 and resides in the same galaxy that hosted
SN candidate AT2016gbz. Likewise, CRTS\_OBC\_324 occurred in the galaxy that
hosted AT2017io. CRTS\_OBC\_331 occurred in the same galaxy as SN 1955Q, and
CRTS\_OBC\_127 occurred in the galaxy that hosted SN 2017me.  The occurrence
of multiple SN within the same host is not unexpected since there are 
many examples of active SN hosts. For example, five SNe were detected 
in NGC 309 over the span of 15 years\footnote{http://www.rochesterastronomy.org/snimages/snhnameall.html}.  

\begin{figure}{
\includegraphics[width=84mm]{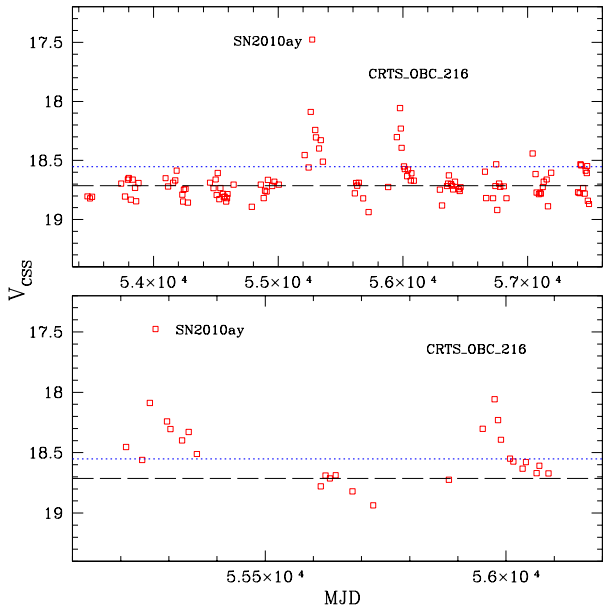}
\caption{\label{CRTS_OBC_216}
Outbursts in SDSS J123527.19+270402.7, including the known 
type-Ic supernova SN 2010ay and the newly discovered 
outburst CRTS\_OBC\_216.
}
}
\end{figure}

One of the more interesting outburst candidates is CRTS\_OBC\_216 (Figure
\ref{CRTS_OBC_216}).  This event occurred in the same host galaxy as known
SN 2010ay.  SN 2010ay was a rare broad line type-Ic SN discovered by CRTS
(Drake et al.~2010). Such SN have long been known to be associated with
long-timescale Gamma Ray Bursts (GRBs, Galama et al.~1998).  In this case,
the host galaxy has low metallicity and exhibits rapid star formation. These
features have also been observed for many GRB-associated SN (Modjaz et
al. 2010).  However, SN 2010ay was noteworthy for lacking any radio or
$\gamma$-ray emission (Sanders et al.~2012).
The CSS lightcurve shows that SN 2010ay is seen to peak in CSS data 
around UT 2010 March 17, while CRTS\_OBC\_216 occurred just two years 
later. Surprisingly, this second event was not noted by transient
surveys or groups performing follow-up of SN 2010ay. 

Inspection of the SDSS images for this galaxy shows minimal extension.  
The locations of the two events are not resolved from each other,
nor from the core of the host in CSS images. Thus, it is not clear 
whether or not the two events are related.

\section{Discussion}

\subsection{Supernovae as a source of outbursts}

The new outburst candidates that we have identified are predominantly
consistent with SNe based on their lightcurve shapes and timescales. In some
cases this is confirmed by the presence of a new source within the
images.  The host galaxies of the outburst candidates are predominantly
starforming galaxies. This latter feature also matches the hosts of most
previously discovered SN candidates in this same sample.  Very few of the
new outbursts are detected in galaxies with known AGN, suggesting few are
likely to be associated with variable black hole accretion.

In comparison to this work, Graur et al.~(2015) found 91 SN in the SDSS
spectra of 740,000 galaxies. Among their detections only 23 SN were new
discoveries. Additionally, Drake et al.~(2014) serendipitously discovered
42 likely SNe from lightcurves while searching for periodic variables in
CSS data. Of those, only six were in galaxies having SDSS spectra.  
Assuming that the fraction new SNe in galaxies from Drake et al.~(2014) 
is representative of the fraction of the full sample, we expect that there 
should be roughly six times as many outbursts ($\sim$4,200) in CSS galaxies 
photometry without SDSS spectra.  
This number is close to the total number of SN candidates that have been
discovered by CRTS. Thus, a very large number of additional outbursts should
be present within the CSS data. Additionally, it is likely that the number 
of detections could be increased by using image subtraction.

Considering what could be learnt from our new outburst candidates, we note
that Kelly \& Kirshner (2012) studied a sample of 519 nearby CCSNe with SDSS
spectra.  They suggested that there was evidence for differences in the
metallicity, the rate of star formation and host offset between the
different types of CCSNe. Such an analysis is not possible with most of the
new outbursts since in most cases the events are not well enough sampled
for type classification. However, the presence of these new likely-SNe in
galaxies shows that the Kelly \& Kirshner (2012) sample was
incomplete. The events missing from the Kelly \& Kirshner (2012) could well
change their conclusions. For example, since most events are not clearly 
offset from their hosts in CSS images, we expect many of the new events are
near the cores of their hosts. These events could clearly affect the
distribution of SN offsets from their hosts.  

Our discovery of new likely-SNe near galaxy nuclei is not surprising. Past 
SN surveys are well known to miss SNe occurring within the cores of
galaxies. The main reasons for this are the higher detection threshold
required (due to surface brightness generally being higher), and the
expectation that events occurring in galaxy cores could be due to 
AGN activity. This AGN bias is compounded by the increased difficulty
of extracting faint SN spectra in the vicinity of a bright nuclear core.

Although our analysis has selected hundreds of new outburst candidates 
it is quite likely that additional candidates could be found via a 
model fitting process~--for example, by searching for events fitting
SN templates. However, as Li et al.~(2011a) note, no type of SN 
can be well represented by a single lightcurve. Thus, the combination
of poorly sampled lightcurves, and a lack of colour information means
that template fitting is unlikely to constrain the types for most 
of the candidates.

\subsection{AGN as an outburst source}

In our current analysis we specifically chose an outburst search 
technique that was model-free in order to minimize assumptions 
about the form the outbursts might take. 
In comparison to recent outbursts found in the Catalina lightcurves 
of QSOs (Graham et al.~2017), the hosts of our outbursts generally 
do not exhibit underlying variability. This makes the interpretation 
of our outbursts clearer than in the case of those with AGN.
Furthermore, this enables us to use a much lower effective
detection threshold than Graham et al.~(2017), or the searches 
of Lawrence et al.~(2016) and Kankare et al.~(2017). 

With this in mind, it is interesting to note that we only find 
evidence for large-amplitude long-timescale flares in a small 
number of galaxies that clearly include AGN.
The general lack of such events in quiescent galaxies strongly 
suggests that the events that have been found in AGN are in 
some way associated with the presence of a SMBH and associated 
accretion disk. 
Nevertheless, we cannot strongly constrain whether the AGN-related 
events in Lawrence et al.~(2016) and Graham et al.(2017) are
in fact lensed by sources in foreground galaxies.
The results do suggest that, if the flares seen in AGN are associated 
with luminous SNe, as suggested for CSS100217 (Drake et al.~2012) and 
PS16dtm (Blanchard et al.~2017) which both reside in NLS1 galaxies, 
they could well be associated with the black hole accretion disk.

\subsection{TDEs as an outburst source}

With the large number of new outburst candidates, it is worth considering
the possibility that some of the events are associated with TDEs. TDEs were
originally predicted to characteristically decline with time as $t^{5/3}$
(Evans \& Kochanek 1989).  However, more recently, widely varying estimates
of the decline rate have been suggested (for example, see Lodato \&
Rossi~2011).  The current data for TDE candidates show that they do indeed
exhibit a wide range of decline timescales (Auchettl et al.~2017). Thus,
the sparsely sampled CSS lightcurves are far from ideal for separating TDEs
with poorly constrained decline rates from SNe that are also well known to
have diverse decline rates.

Considering other methods for identifying TDEs, Hung et al.~(2017) recently
undertook an analysis of iPTF data observed as part of their TDE search.  In
their work they suggested that high precision astrometry is essential for
finding TDEs, their reasoning being that type-Ia SNe have similar colours
and are far more common than TDEs (even within the cores of galaxies,
i.e. $< 0.8\arcsec$ from the center). However, we note that a truly
definitive separation requires optical spectra and ideally an X-ray
signature. The spectroscopic signature of TDEs is predominately 
considered to be the presence of strong, broad helium lines (Gezari et al.~2012).

In their survey, Hung et al.~(2017) required outbursts reaching a threshold
of $\Delta M > 0.5$ mag. This is a higher threshold than many of the events
we have found in this analysis. We note that both our work and the Hung et
al.~(2017) TDE survey are necessarily biased against the discovery of events
in intrinsically bright galaxies. 

Unlike our current analysis, Hung et al.~(2017) concentrated their 
analysis solely on red galaxies. They reasoned that most known TDEs have 
so far been found is such galaxies. However, it seems likely that this 
is at least partly due to the fact that prior searches included a strong 
selection bias against other types of galaxies. That is to say, like SN 
surveys, many past searches for TDEs specifically avoid galaxies where 
an AGN might be present due to the difficulty of disentangling the 
signature of a TDE from an AGN. Thus galaxies with blue continua 
(due to the presence of AGN or star formation), are under 
represented in the search samples. 
Indeed, Blanchard et al.~(2017) suggest that AGN flares may be due to TDEs,
and that these may be more common in AGN than previously believed. In
addition, Tadhunter et al.~(2017) suggest that the rate of TDEs in
ultra-luminous infrared galaxies may be much higher than in normal galaxies.
Indeed, the numerous flaring events that have recently been found within AGN
(Drake et al.~2011, Lawrence et al.~2016, Graham et al.~2017, Blanchard et
al.~2017) appear to be roughly consistent with a TDE origin. On the other hand, 
Lawrence et al.~(2016) suggests, the amount of energy released by a 
TDE is far less than seen in many of the observed AGN flaring events.

Overall, it is possible that some of the new outbursts are due to TDEs.
In particular, CRTS\_OBC\_204 is clearly not a regular SN, and occurred 
within a host containing an AGN. However, since predicted TDE rates are 
so much lower than SN rates, we do not expect many of the outbursts 
to be due to TDEs.

\subsection{Considerations for the future}

As we look forward to larger transient surveys that are coming
(such as ZTF and LSST), we expect an ever growing number of 
transient discoveries. As with current surveys, a very large 
fraction of the transients will undoubtedly be SNe. It is 
therefore worth considering what may be learnt from such 
objects when there are already far more SN being discovered 
than can be spectroscopically confirmed.

\begin{figure}{
\includegraphics[width=84mm]{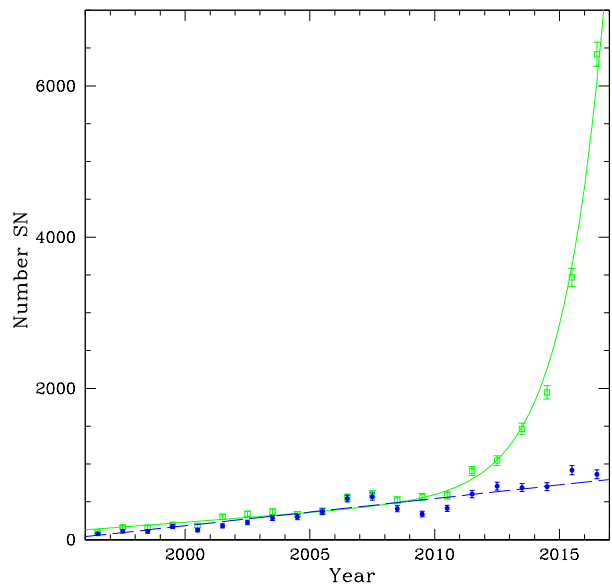}
\caption{\label{SNdates}
The number of yearly SN candidate detections 
is plotted as green points and solid line. The number 
of spectroscopically confirmed SN are plotted as the 
dashed-blue line and points. The lines are included 
only to show the apparent trends.
}
}
\end{figure}

In Figure \ref{SNdates}, we plot the number of discovered 
SN candidates and the number of confirmed SN over the
past 22 years. Encouragingly, this figure shows that the amount 
of SN follow-up has increased at an approximately linear 
rate over this time. However, in contrast, the SN candidate 
discovery rate appears well modeled by an exponential 
increase in addition to the linear confirmation-rate trend. 

Examining the SN confirmation data we find that the shortfall in SN confirmation
is highly magnitude dependent.  Specifically, based on historical SN data,
only 10 to 20\% of the SN discovered since 1950 with $20 < V < 22$ have been
spectroscopically confirmed, while roughly 80\% of SN
candidates brighter than $V=17$ have been spectroscopically confirmed.

In the near term, surveys such as ZTF and LSST will almost certainly
increase the SN discovery trend by introducing many more transients
with luminosities that have not been historically followed. The very faint
events discovered by LSST will be particularly difficult to confirm
spectroscopically. Characterisation will be further hindered by the
fact that LSST lightcurves will not be as well sampled as many existing
transient surveys.  For brighter events, it is possible to mitigate the {\em
follow-up problem} to some extent by using historical information. However,
a better understanding of fainter events should also make use of the data
for the tens of thousands of brighter transients that have been already been
found by existing transient surveys.  These should be used to glean a fuller
understanding of the event parameter space.

As new transients are discovered in future surveys, the existence of better 
sampled lightcurves and colour information (in the case of ZTF), combined 
with accurate astrometry (from {\it Gaia}), will undoubtedly aid our ability
to classify optical transients. Nevertheless, a significant amount of 
additional characterization work is required now since spectroscopic 
follow-up resources are unlikely to have a significant impact on 
the overall fraction of characterised events. Classification and follow-up 
also remain crucial for determining the nature of long timescale AGN-associated events 
as well as other very rare types of transients such as GW170817 (Abbott et al.~2017) 
and 2018cow (Perley et al.~2018).

\section{Summary}

In this work we have taken a step towards understanding the types of
extra galactic transients that were missed by existing transient surveys.
We used lightcurves from the Catalina Surveys to find them, and archival 
spectra to help characterise their hosts. 

The main discovery of this work has been the discovery of more than 700 new
outbursts among more than 1.4 million spectroscopically confirmed SDSS
galaxies. As expected, we find that most of these events are consistent with
SNe. However, we cannot exclude the presence of rare transient types since
specific classifications are only possible for a small number of the
sources. Nevertheless, we find a small number of long timescale events that
are associated with galaxies containing AGN. These outbursts appear
relatively consistent with the events discovered in AGN-containing galaxies
by Lawrence et al.~(2016) and Graham et al.~(2017). Importantly, we do not
find these long-timescale events among the $>95\%$ of galaxies in our sample
that lack a significant evidence of AGN. This strongly suggests that the
long timescale events are in some way associated with AGN. As yet we are
unable to determine whether the source of the long outbursts are intrinsic
to AGN or their environment, or are due to an extrinsic effect such as
microlensing of the AGN.

\section*{Acknowledgments}

CRTS, CSDR1 and CSDR2 are supported by the U.S.~National Science Foundation under
grant NSF grants AST-1313422, AST-1413600, and AST-1518308.  The CSS survey
is funded by the National Aeronautics and Space Administration under Grant
No. NNG05GF22G issued through the Science Mission Directorate Near-Earth Objects
Observations Program. AJD and MC acknowledge partial support by CONICYT's 
PCI program through grant DPI20140066. MC is additionally supported by the 
Ministry for the Economy, Development, and Tourism's Iniciativa Cient\'{i}fica 
Milenio through grant IC\,120009, awarded to the Millennium Institute of Astrophysics; 
by Proyecto Fondecyt Regular \#1171273; and by Proyecto Basal PFB-06/2007. 
AllWISE makes use of data from WISE, which is a joint
project of the University of California, Los Angeles, and the Jet
Propulsion Laboratory/California Institute of Technology, and NEOWISE,
which is a project of the Jet Propulsion Laboratory/California
Institute of Technology. WISE and NEOWISE are funded by the 
National Aeronautics and Space Administration.

\clearpage

\end{document}